\documentclass[italian,english]{article}
\usepackage[T1]{fontenc}
\usepackage[latin1]{inputenc}
\usepackage{color}
\usepackage{graphicx}
\usepackage{amssymb}

\makeatletter

 \newcommand{\lyxaddress}[1]{
   \par {\raggedright #1 
   \vspace{1.4em}
   \noindent\par}
 }

\usepackage{babel}
\makeatother
\begin{document}

\title{\textbf{''Magnetic'' components of gravitational waves and response
functions of interferometers}}

\author{\textbf{Christian Corda}}

\maketitle

\lyxaddress{\begin{center}Associazione Scientifica Galileo Galilei, Via Pier
Cironi 16 - 59100 PRATO, Italy; Dipartimento della Gravitazione, Centro
Scienze Naturali, via di Galceti 74 - 59100 PRATO, Italy\end{center}}

\lyxaddress{\begin{center}\textit{E-mail address:} \textcolor{blue}{christian.corda@ego-gw.it} \end{center}}

\begin{abstract}
Recently, arising from an enlighting analysis of Baskaran and Grishchuk
in \textit{Class. Quant. Grav.} \textbf{\textit{21}} \textit{4041-4061
(2004)}, some papers in the literature have shown the presence and
importance of the so-called {}``magnetic'' components of gravitational
waves (GWs), which have to be taken into account in the context of
the total response functions of interferometers for GWs propagating
from arbitrary directions. In \foreignlanguage{italian}{\textit{Int.
Journ.}} \textit{Mod. Phys. A 22, 13, 2361-2381 (2007)} and \textit{Int.
J. Mod. Phys. D} \textbf{\textit{16,}} \textit{9, 1497-1517  (2007)}
accurate response functions for the Virgo and LIGO interferometers
have been analysed. 

However, some results which have been shown in \foreignlanguage{italian}{\textit{Int.
Journ.}} \textit{Mod. Phys. A 22, 13, 2361-2381 (2007)} look in contrast
with the results which have been shown in \textit{Int. J. Mod. Phys.
D} \textbf{\textit{16,}} \textit{9, 1497-1517  (2007)}. In fact, in
\foreignlanguage{italian}{\textit{Int. Journ.}} \textit{Mod. Phys.
A 22, 13, 2361-2381 (2007)} it was claimed that the {}``magnetic''
component of GWs could, in principle, extend the frequency range of
Earth based interferometers, while in \textit{Int. J. Mod. Phys. D}
\textbf{\textit{16,}} \textit{9, 1497-1517  (2007)} such a possibility
has been banned.

This contrast has been partially solved in the \textit{Proceedings
of the XLIInd Rencontres de Moriond, Gravitational Waves and Experimental
Gravity, La Thuile, Val d'Aosta Italy (March 12-18 2007).}

The aim of this review paper is to re-analyse all the framework of
the {}``magnetic'' components of GWs with the goal of solving the
mentioned contrast in definitive way.

Accurate response funtions for the Virgo and LIGO interferometers
will be also re-discussed in detail.
\end{abstract}

\lyxaddress{PACS numbers: 04.80.Nn, 04.80.-y, 04.25.Nx}

\section{Introduction}

The data analysis of interferometric gravitational waves (GWs) detectors
has recently been started (for the current status of GWs interferometers
see \cite{key-1,key-2,key-3,key-4,key-5,key-6,key-7,key-8}) and the
scientific community aims in a first direct detection of GWs in next
years. 

Detectors for GWs will be important for a better knowledge of the
Universe and also to confirm or ruling out the physical consistency
of General Relativity or of any other theory of gravitation \cite{key-9,key-10,key-11,key-12,key-13,key-14,key-15,key-16}.
This is because, in the context of Extended Theories of Gravity, some
differences between General Relativity and the others theories can
be pointed out starting by the linearized theory of gravity \cite{key-9,key-10,key-12,key-14}.
In this picture, detectors for GWs are in principle sensitive also
to a hypotetical \textit{scalar} component of gravitational radiation,
that appears in extended theories of gravity like scalar-tensor gravity,
high order theories \cite{key-12,key-15,key-16,key-17,key-18,key-19,key-20,key-21,key-22}
and Brans-Dicke theory \cite{key-23}.

Recently, arising from an enlighting analysis of Baskaran and Grishchuk
in \cite{key-24}, some papers in the literature have shown the presence
and importance of the so-called {}``magnetic'' components of GWs,
which have to be taken into account in the context of the total response
functions of interferometers for GWs propagating from arbitrary directions.
In \cite{key-25} and \cite{key-26} accurate response functions for
the Virgo and LIGO interferometers have been analysed. 

However, some results which have been shown in \cite{key-25} look
in contrast with the results which have been shown in\cite{key-26}.
In fact, in \cite{key-25} it was claimed that the {}``magnetic''
component of GWs could, in principle, extend the frequency range of
Earth based interferometers, while in \cite{key-26} such a possibility
has been banned.

This contrast has been partially solved in \cite{key-27}\textit{.}

The aim of this review paper is to re-analyse all the framework of
the {}``magnetic'' components of GWs with the goal of solving the
mentioned contrast in definitive way.

Accurate response funtions for the Virgo and LIGO interferometers
will be re-discussed in detail too.

\section{Analysis in the frame of the local observer}

In a laboratory enviroment on earth, the coordinate system in which
the space-time is locally flat is typically used and the distance
between any two points is given simply by the difference in their
coordinates in the sense of Newtonian physics \cite{key-9,key-12,key-24,key-25,key-26,key-27,key-28}.
In this frame, called the frame of the local observer, GWs manifest
themself by exerting tidal forces on the masses (the mirror and the
beam-splitter in the case of an interferometer, see figure 1). 

\begin{figure}
\includegraphics{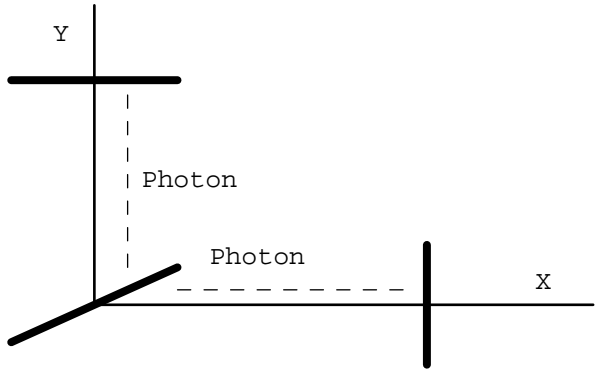}

\caption{photons can be launched from the beam-splitter to be bounced back
by the mirror}
\end{figure}

Recently, the presence and importance of the so-called magnetic components
of GWs have been shown by Baskaran and Grishchuk that computed the
correspondent detector patterns in the low-frequencies approximation
\cite{key-24}. Then, more detailed angular and frequency dependences
of the response functions for the magnetic components has been given
in the same approximation, with a specific application to the parameters
of the LIGO and Virgo interferometers in \cite{key-25,key-26,key-27}.
The most important goal of this review paper is to solve in a definitive
way the contrast between \cite{key-25} and \cite{key-26} on the
possibility of extending the frequency-band of interferometers.

Before starting with the analysis of the response functions, a brief
review of Section 3 of \cite{key-24} is due to understand the importance
of the {}``magnetic'' components of GWs. In this review paper we
will use different notations with respect to the ones used in \cite{key-24}.
Following \cite{key-25,key-26,key-27}, we work with $G=1$, $c=1$
and $\hbar=1$ and we call $h_{+}(t_{tt}+z_{tt})$ and $h_{\times}(t_{tt}+z_{tt})$
the weak perturbations due to the $+$ and the $\times$ polarizations
which are expressed in terms of synchronous coordinates $t_{tt},x_{tt},y_{tt},z_{tt}$
in the transverse-traceless (TT) gauge. In this way, the most general
GW propagating in the $z_{tt}$ direction can be written in terms
of plane monochromatic waves \cite{key-25,key-26,key-27}

\begin{equation}
\begin{array}{c}
h_{\mu\nu}(t_{tt}+z_{tt})=h_{+}(t_{tt}+z_{tt})e_{\mu\nu}^{(+)}+h_{\times}(t_{tt}+z_{tt})e_{\mu\nu}^{(\times)}=\\
\\=h_{+0}\exp i\omega(t_{tt}+z_{tt})e_{\mu\nu}^{(+)}+h_{\times0}\exp i\omega(t_{tt}+z_{tt})e_{\mu\nu}^{(\times)},\end{array}\label{eq: onda generale}\end{equation}

and the correspondent line element will be

\begin{equation}
ds^{2}=dt_{tt}^{2}-dz_{tt}^{2}-(1+h_{+})dx_{tt}^{2}-(1-h_{+})dy_{tt}^{2}-2h_{\times}dx_{tt}dx_{tt}.\label{eq: metrica TT totale}\end{equation}

The wordlines $x_{tt},y_{tt},z_{tt}=const.$ are timelike geodesics
representing the histories of free test masses \cite{key-24,key-25,key-26,key-27}.
The coordinate transformation $x^{\alpha}=x^{\alpha}(x_{tt}^{\beta})$
from the TT coordinates to the frame of the local observer is \cite{key-24,key-25,key-26,key-27}.

\begin{equation}
\begin{array}{c}
t=t_{tt}+\frac{1}{4}(x_{tt}^{2}-y_{tt}^{2})\dot{h}_{+}-\frac{1}{2}x_{tt}y_{tt}\dot{h}_{\times}\\
\\x=x_{tt}+\frac{1}{2}x_{tt}h_{+}-\frac{1}{2}y_{tt}h_{\times}+\frac{1}{2}x_{tt}z_{tt}\dot{h}_{+}-\frac{1}{2}y_{tt}z_{tt}\dot{h}_{\times}\\
\\y=y_{tt}+\frac{1}{2}y_{tt}h_{+}-\frac{1}{2}x_{tt}h_{\times}+\frac{1}{2}y_{tt}z_{tt}\dot{h}_{+}-\frac{1}{2}x_{tt}z_{tt}\dot{h}_{\times}\\
\\z=z_{tt}-\frac{1}{4}(x_{tt}^{2}-y_{tt}^{2})\dot{h}_{+}+\frac{1}{2}x_{tt}y_{tt}\dot{h}_{\times},\end{array}\label{eq: trasf. coord.}\end{equation}

where it is $\dot{h}_{+}\equiv\frac{\partial h_{+}}{\partial t}$
and $\dot{h}_{\times}\equiv\frac{\partial h_{\times}}{\partial t}$.
The coefficients of this transformation (components of the metric
and its first time derivative) are taken along the central wordline
of the local observer \cite{key-24,key-25,key-26,key-27}. It is well
known from \cite{key-24,key-25,key-26,key-27} that the linear and
quadratic terms, as powers of $x_{tt}^{\alpha}$, are unambiguously
determined by the conditions of the frame of the local observer, while
the cubic and higher-order corrections are not determined by these
conditions. Thus, at high-frequencies, the expansion in terms of higher-order
corrections breaks down \cite{key-24,key-26,key-27}.

Considering a free mass riding on a timelike geodesic ($x=l_{1}$,
$y=l_{2},$ $z=l_{3}$) \cite{key-24,key-25,key-26,key-27}, eqs.
(\ref{eq: trasf. coord.}) define the motion of this mass with respect
to the introduced frame of the local observer. In concrete terms one
gets\begin{equation}
\begin{array}{c}
x(t)=l_{1}+\frac{1}{2}[l_{1}h_{+}(t)-l_{2}h_{\times}(t)]+\frac{1}{2}l_{1}l_{3}\dot{h}_{+}(t)+\frac{1}{2}l_{2}l_{3}\dot{h}_{\times}(t)\\
\\y(t)=l_{2}-\frac{1}{2}[l_{2}h_{+}(t)+l_{1}h_{\times}(t)]-\frac{1}{2}l_{2}l_{3}\dot{h}_{+}(t)+\frac{1}{2}l_{1}l_{3}\dot{h}_{\times}(t)\\
\\z(t)=l_{3}-\frac{1}{4[}(l_{1}^{2}-l_{2}^{2})\dot{h}_{+}(t)+2l_{1}l_{2}\dot{h}_{\times}(t),\end{array}\label{eq: Grishuk 0}\end{equation}
which are exactly eqs. (13) of \cite{key-24} rewritten using our
notation. In absence of GWs the position of the mass is $(l_{1},l_{2},l_{3}).$
The effect of the GW is to drive the mass to have oscillations. Thus,
in general, from eqs. (\ref{eq: Grishuk 0}) all three components
of motion are present \cite{key-24,key-25,key-26,key-27}.

Neglecting the terms with $\dot{h}_{+}$ and $\dot{h}_{\times}$ in
eqs. (\ref{eq: Grishuk 0}), the {}``traditional'' equations for
the mass motion are obtained \cite{key-24,key-25,key-26,key-27}:\begin{equation}
\begin{array}{c}
x(t)=l_{1}+\frac{1}{2}[l_{1}h_{+}(t)-l_{2}h_{\times}(t)]\\
\\y(t)=l_{2}-\frac{1}{2}[l_{2}h_{+}(t)+l_{1}h_{\times}(t)]\\
\\z(t)=l_{3}.\end{array}\label{eq: traditional}\end{equation}

Clearly, this is the analogous of the electric component of motion
in electrodynamics \cite{key-24,key-25,key-26,key-27}, while equations\begin{equation}
\begin{array}{c}
x(t)=l_{1}+\frac{1}{2}l_{1}l_{3}\dot{h}_{+}(t)+\frac{1}{2}l_{2}l_{3}\dot{h}_{\times}(t)\\
\\y(t)=l_{2}-\frac{1}{2}l_{2}l_{3}\dot{h}_{+}(t)+\frac{1}{2}l_{1}l_{3}\dot{h}_{\times}(t)\\
\\z(t)=l_{3}-\frac{1}{4[}(l_{1}^{2}-l_{2}^{2})\dot{h}_{+}(t)+2l_{1}l_{2}\dot{h}_{\times}(t),\end{array}\label{eq: news}\end{equation}

are the analogous of the magnetic component of motion. One could think
that the presence of these magnetic components is a {}``frame artefact''
due to the transformation (\ref{eq: trasf. coord.}), but in Section
4 of \cite{key-24} eqs. (\ref{eq: Grishuk 0}) have been directly
obtained from the geodesic deviation equation too, thus the magnetic
components have a real physical significance. The fundamental point
of \cite{key-24,key-26,key-27} is that the magnetic components become
important when the frequency of the wave increases (Section 3 of \cite{key-24}),
but only in the low-frequency regime. This can be understood directly
from eqs. (\ref{eq: Grishuk 0}). In fact, using eqs. (\ref{eq: onda generale})
and (\ref{eq: trasf. coord.}), eqs. (\ref{eq: Grishuk 0}) become\begin{equation}
\begin{array}{c}
x(t)=l_{1}+\frac{1}{2}[l_{1}h_{+}(t)-l_{2}h_{\times}(t)]+\frac{1}{2}l_{1}l_{3}\omega h_{+}(t-\frac{\pi}{2})+\frac{1}{2}l_{2}l_{3}\omega h_{\times}(t-\frac{\pi}{2})\\
\\y(t)=l_{2}-\frac{1}{2}[l_{2}h_{+}(t)+l_{1}h_{\times}(t)]-\frac{1}{2}l_{2}l_{3}\omega h_{+}(t-\frac{\pi}{2})+\frac{1}{2}l_{1}l_{3}\omega h_{\times}(t-\frac{\pi}{2})\\
\\z(t)=l_{3}-\frac{1}{4[}(l_{1}^{2}-l_{2}^{2})\omega h_{+}(t-\frac{\pi}{2})+2l_{1}l_{2}\omega h_{\times}(t-\frac{\pi}{2}).\end{array}\label{eq: Grishuk 01}\end{equation}

Thus, the terms with $\dot{h}_{+}$ and $\dot{h}_{\times}$ in eqs.
(\ref{eq: Grishuk 0}) can be neglected only when the wavelength goes
to infinity \cite{key-24,key-25,key-26,key-27}, while, at high-frequencies,
the expansion in terms of $\omega l_{i}l_{j}$ corrections, with $i,j=1,2,3,$
breaks down \cite{key-24,key-26,key-27}. This fact has not been emphasized
in \cite{key-25}, thus one could think that the {}``magnetic''
comonents of GWs could, in principle, extend the frequency-range of
interferometers, but this is not correct \cite{key-24,key-25,key-26,key-27}.

Now, let us compute the total response functions of interferometers
for the magnetic components.

Equations (\ref{eq: Grishuk 0}), that represent the coordinates of
the mirror of the interferometer in presence of a GW in the frame
of the local observer, can be rewritten for the pure magnetic component
of the $+$ polarization as

\begin{equation}
\begin{array}{c}
x(t)=l_{1}+\frac{1}{2}l_{1}l_{3}\dot{h}_{+}(t)\\
\\y(t)=l_{2}-\frac{1}{2}l_{2}l_{3}\dot{h}_{+}(t)\\
\\z(t)=l_{3}-\frac{1}{4}(l_{1}^{2}-l_{2}^{2})\dot{h}_{+}(t),\end{array}\label{eq: Grishuk 1}\end{equation}

where $l_{1},l_{2}\textrm{ }and\textrm{ }\textrm{ }l_{3}$ are the
unperturbed coordinates of the mirror. 

To compute the response functions for an arbitrary propagating direction
of the GW, we recall that the arms of the interferometer are in general
in the $\overrightarrow{u}$ and $\overrightarrow{v}$ directions,
while the $x,y,z$ frame is adapted to the propagating GW (i.e. the
observer is assumed located in the position of the beam splitter).
Then, a spatial rotation of the coordinate system has to be performed:

\begin{equation}
\begin{array}{ccc}
u & = & -x\cos\theta\cos\phi+y\sin\phi+z\sin\theta\cos\phi\\
\\v & = & -x\cos\theta\sin\phi-y\cos\phi+z\sin\theta\sin\phi\\
\\w & = & x\sin\theta+z\cos\theta,\end{array}\label{eq: rotazione magn}\end{equation}

or, in terms of the $x,y,z$ frame:

\begin{equation}
\begin{array}{ccc}
x & = & -u\cos\theta\cos\phi-v\cos\theta\sin\phi+w\sin\theta\\
\\y & = & u\sin\phi-v\cos\phi\\
\\z & = & u\sin\theta\cos\phi+v\sin\theta\sin\phi+w\cos\theta.\end{array}\label{eq: rotazione 2 magn}\end{equation}

In this way the GW is propagating from an arbitrary direction $\overrightarrow{r}$
to the interferometer (see figure 2 ). 

\begin{figure}
\includegraphics{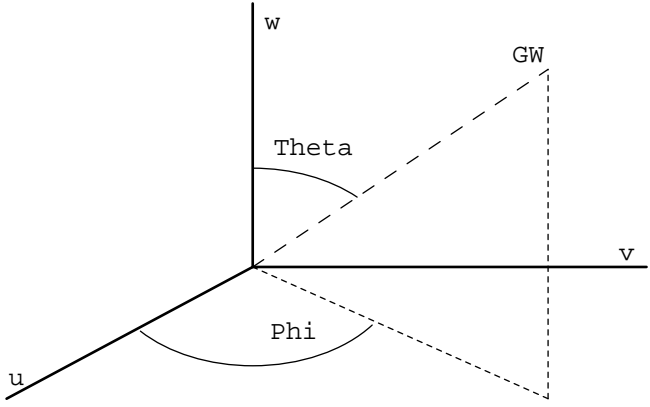}

\caption{a GW propagating from an arbitrary direction}
\end{figure}
As the mirror of eqs. (\ref{eq: Grishuk 1}) is situated in the $u$
direction, using eqs. (\ref{eq: Grishuk 1}), (\ref{eq: rotazione magn})
and (\ref{eq: rotazione 2 magn}) the $u$ coordinate of the mirror
is given by

\begin{equation}
u=L+\frac{1}{4}L^{2}A\dot{h}_{+}(t),\label{eq: du magn}\end{equation}

where \begin{equation}
A\equiv\sin\theta\cos\phi(\cos^{2}\theta\cos^{2}\phi-\sin^{2}\phi)\label{eq: A}\end{equation}

and $L=\sqrt{l_{1}^{2}+l_{2}^{2}+l_{3}^{2}}$ is the length of the
interferometer arms.

The computation for the $v$ arm is similar to the one above. Using
eqs. (\ref{eq: Grishuk 1}), (\ref{eq: rotazione magn}) and (\ref{eq: rotazione 2 magn}),
the coordinate of the mirror in the $v$ arm is:

\begin{equation}
v=L+\frac{1}{4}L^{2}B\dot{h}_{+}(t),\label{eq: dv magn}\end{equation}

where\begin{equation}
B\equiv\sin\theta\sin\phi(\cos^{2}\theta\cos^{2}\phi-\sin^{2}\phi).\label{eq: B}\end{equation}

\section{The response function of an interferometer for the magnetic contribution
of the $+$ polarization}

Equations (\ref{eq: du magn}) and (\ref{eq: dv magn}) represent
the distance of the two mirrors of the interferometer from the beam-splitter
in presence of the GW (note that only the contribution of the magnetic
component of the $+$ polarization of the GW is taken into account).
They represent particular cases of the more general form given in
eq. (33) of \cite{key-24}.

A {}``signal'' can also be defined in the time domain ($T=L$ in
our notation):

\begin{equation}
\frac{\delta T(t)}{T}\equiv\frac{u-v}{L}=\frac{1}{4}L(A-B)\dot{h}_{+}(t).\label{eq: signal piu}\end{equation}

The quantity (\ref{eq: signal piu}) can be computed in the frequency
domain using the Fourier transform of $h_{+}$, defined by

\begin{equation}
\tilde{h}_{+}(\omega)=\int_{-\infty}^{\infty}dth_{+}(t)\exp(i\omega t),\label{eq: trasformata di fourier magn}\end{equation}
obtaining

\[
\frac{\tilde{\delta}T(\omega)}{T}=H_{magn}^{+}(\omega)\tilde{h}_{+}(\omega),\]

where the function

\begin{equation}
\begin{array}{c}
H_{magn}^{+}(\omega)=-\frac{1}{8}i\omega L(A-B)=\\
\\=-\frac{1}{4}i\omega L\sin\theta[(\cos^{2}\theta+\sin2\phi\frac{1+\cos^{2}\theta}{2})](\cos\phi-\sin\phi)\end{array}\label{eq: risposta totale magn}\end{equation}

is the total response function of the interferometer for the magnetic
component of the $+$ polarization, in perfect agreement with the
result of Baskaran and Grishchuk (eqs. 46 and 49 of \cite{key-24}).
In the above computation the theorem on the derivative of the Fourier
transform has been used.

In the present work the $x,y,z$ frame is the frame of the local observer
adapted to the propagating GW, while in \cite{key-24} the two frames
are not in phase (i.e. in this paper the third angle is put equal
to zero, this is not a restriction as it is known in the literature
\cite{key-25,key-26,key-27}).

The absolute value of the response functions (\ref{eq: risposta totale magn})
of the Virgo ($L=3$Km) and LIGO ($L=4$Km) interferometers to the
magnetic component of the $+$ polarization for $\theta=\frac{\pi}{4}$
and $\phi=\frac{\pi}{3}$ are respectively shown in figures 3 and
4 in the low-frequency range $10Hz\leq f\leq100Hz$. This quantity
increases with increasing frequency.%
\begin{figure}
\includegraphics{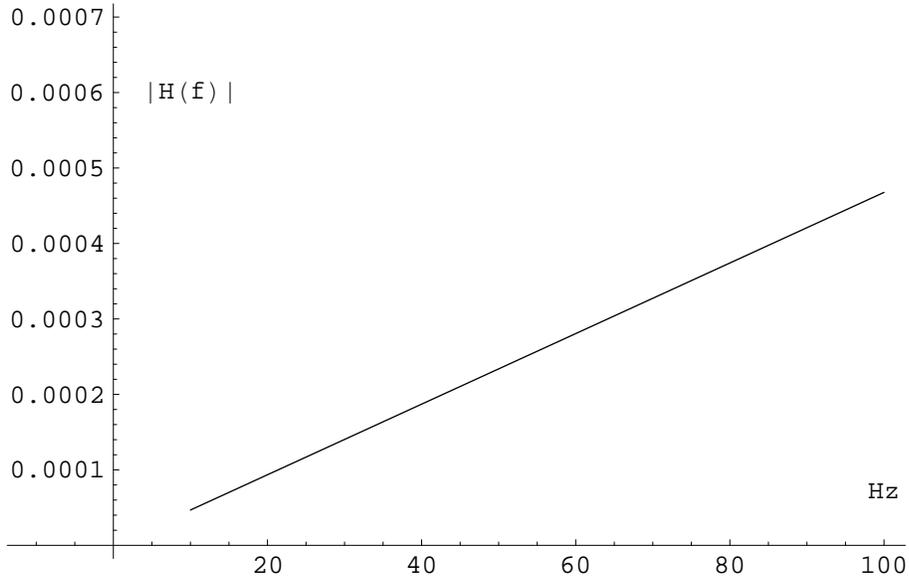}

\caption{the absolute value of the total response function of the Virgo interferometer
to the magnetic component of the $+$ polarization for $\theta=\frac{\pi}{4}$
and $\phi=\frac{\pi}{3}$ in the low-frequency range $10Hz\leq f\leq100Hz$}
\end{figure}
\begin{figure}
\includegraphics{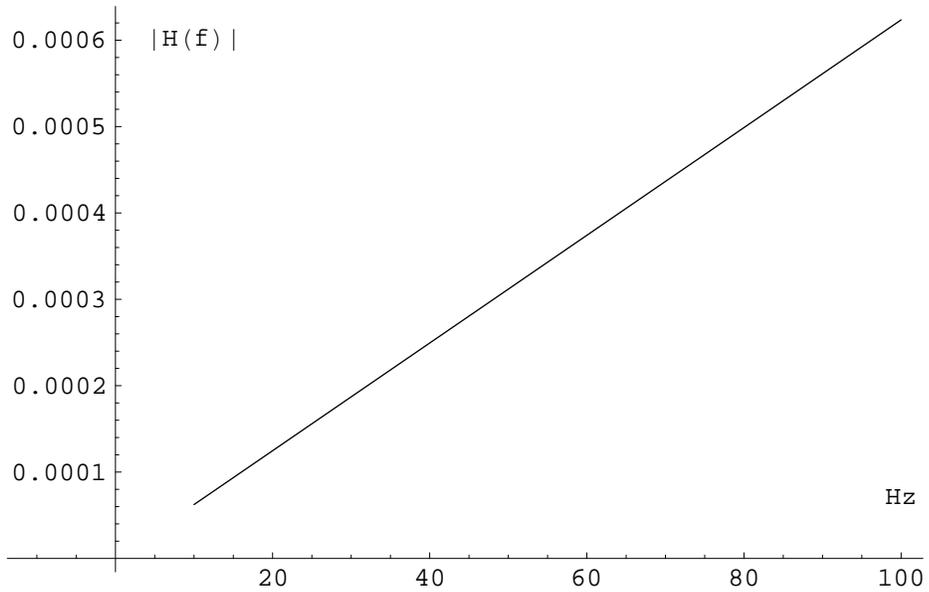}

\caption{the absolute value of the total response function of the LIGO interferometer
to the magnetic component of the $+$ polarization for $\theta=\frac{\pi}{4}$
and $\phi=\frac{\pi}{3}$ in the low- frequency range $10Hz\leq f\leq100Hz$}
\end{figure}
The angular dependences of the response function (\ref{eq: risposta totale magn})
of the Virgo and LIGO interferometers to the magnetic component of
the $+$ polarization for $f=100Hz$ are shown in figures 5 and 6. 

\begin{figure}
\includegraphics{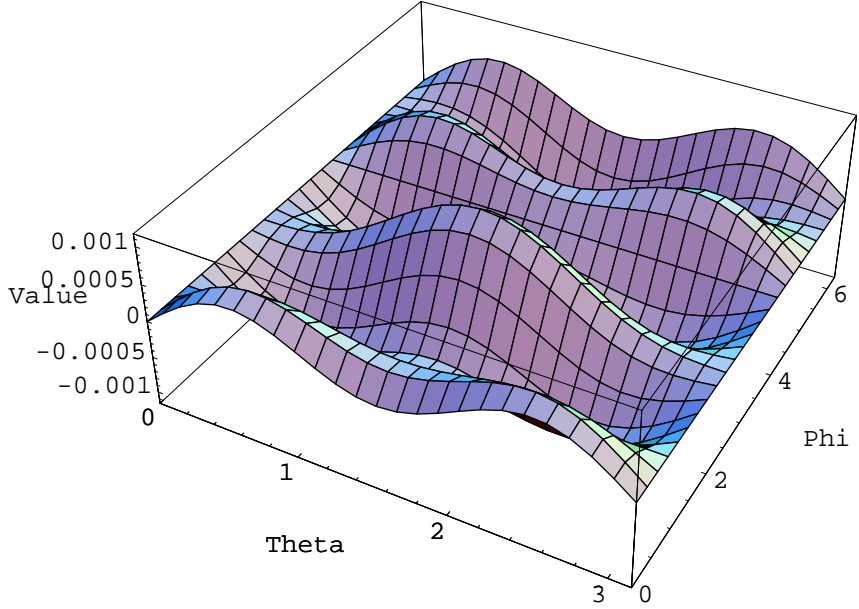}

\caption{the angular dependence of the response function of the Virgo interferometer
to the magnetic component of the $+$ polarization for $f=100Hz$}
\end{figure}
\begin{figure}
\includegraphics{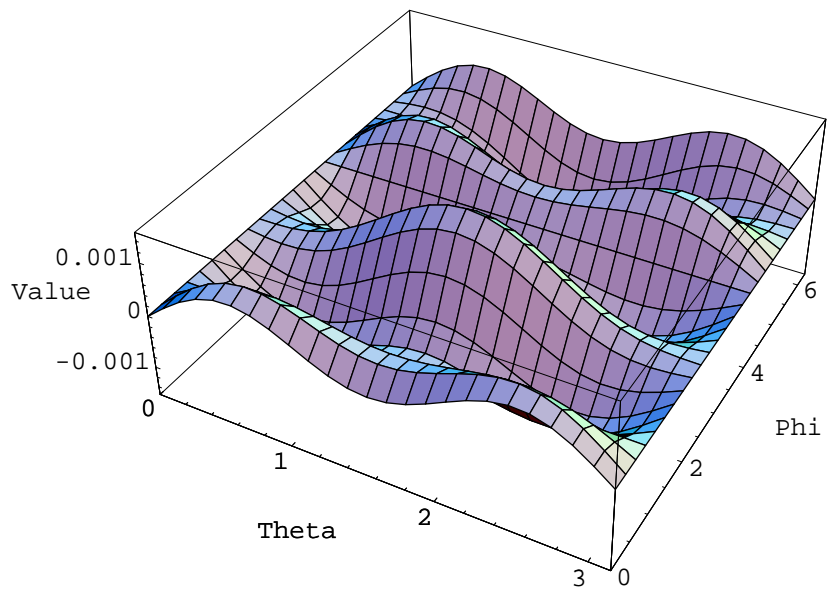}

\caption{the angular dependence of the response function of the LIGO interferometer
to the magnetic component of the $+$ polarization for $f=100Hz$}
\end{figure}

\section{Analysis for the $\times$ polarization}

The analysis can be generalized for the magnetic component of the
$\times$ polarization too. In this case, equations (\ref{eq: Grishuk 0})
can be rewritten for the pure magnetic component of the $\times$
polarization as

\begin{equation}
\begin{array}{c}
x(t)=l_{1}+\frac{1}{2}l_{2}l_{3}\dot{h}_{\times}(t)\\
\\y(t)=l_{2}+\frac{1}{2}l_{1}l_{3}\dot{h}_{\times}(t)\\
\\z(t)=l_{3}-\frac{1}{2}l_{1}l_{2}\dot{h}_{\times}(t).\end{array}\label{eq: Grishuk 2}\end{equation}

Using eqs. (\ref{eq: Grishuk 2}), (\ref{eq: rotazione magn}) and
(\ref{eq: rotazione 2 magn}), the $u$ coordinate of the mirror in
the $u$ arm of the interferometer is given by \begin{equation}
u=L+\frac{1}{4}L^{2}C\dot{h}_{\times}(t),\label{eq: du C}\end{equation}

where \begin{equation}
C\equiv-2\cos\theta\cos^{2}\phi\sin\theta\sin\phi,\label{eq: C}\end{equation}
while the $v$ coordinate of the mirror in the $v$ arm of the interferometer
is given by \begin{equation}
v=L+\frac{1}{4}L^{2}D\dot{h}_{\times}(t),\label{eq: dv  D}\end{equation}

where \begin{equation}
D\equiv2\cos\theta\cos\phi\sin\theta\sin^{2}\phi.\label{eq: D}\end{equation}

Thus, with an analysis similar to the one of previous Sections, it
is possible to show that the response function of the interferometer
for the magnetic component of the $\times$ polarization is\begin{equation}
\begin{array}{c}
H_{magn}^{\times}(\omega)=-i\omega T(C-D)=\\
\\=-i\omega L\sin2\phi(\cos\phi+\sin\phi)\cos\theta,\end{array}\label{eq: risposta totale 2 per magn}\end{equation}

in perfect agreement with the result of Baskaran and Grishchuk (eqs.
46 and 50 of \cite{key-24}). The absolute value of the total response
functions (\ref{eq: risposta totale 2 per magn}) of the Virgo and
LIGO interferometers to the magnetic component of the $\times$ polarization
for $\theta=\frac{\pi}{4}$ and $\phi=\frac{\pi}{3}$ are respectively
shown in figure 7 and 8 in the low- frequency range $10Hz\leq f\leq100Hz$.
This quantity increases with increasing frequency in analogy with
the case shown in previous Section for the magnetic component of the
$+$ polarization. %
\begin{figure}
\includegraphics{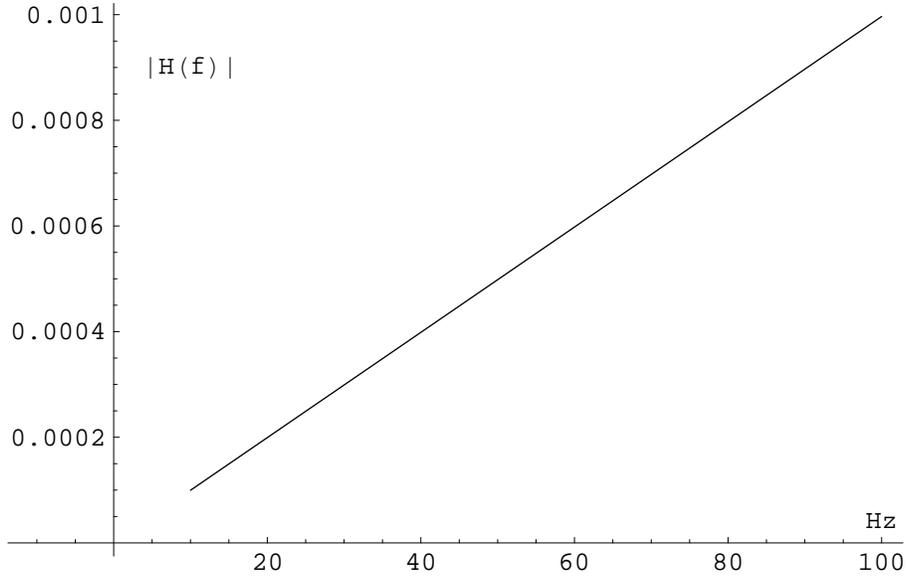}

\caption{the absolute value of the total response function of the Virgo interferometer
to the magnetic component of the $\times$ polarization for $\theta=\frac{\pi}{4}$
and $\phi=\frac{\pi}{3}$ in the low- frequency range $10Hz\leq f\leq100Hz$}
\end{figure}
\begin{figure}
\includegraphics{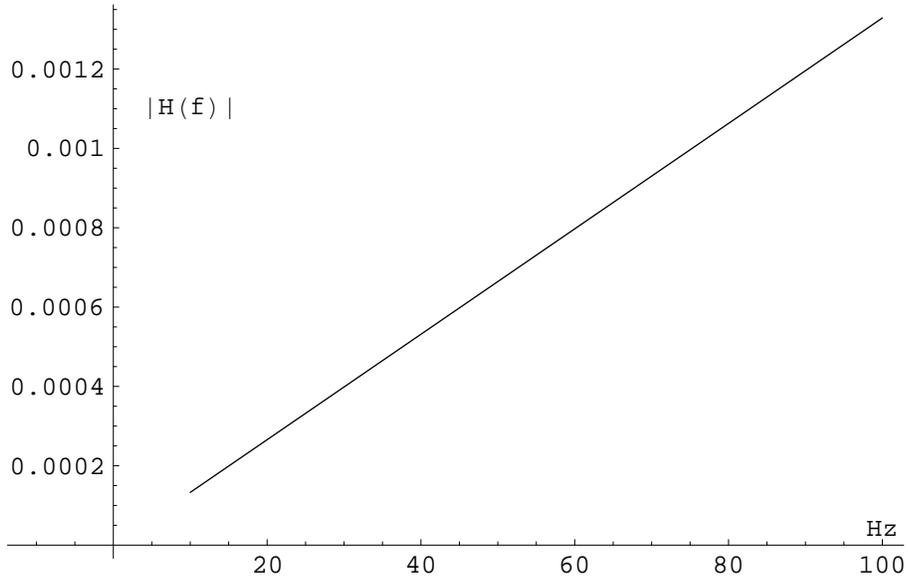}

\caption{the absolute value of the total response function of the LIGO interferometer
to the magnetic component of the $\times$ polarization for $\theta=\frac{\pi}{4}$
and $\phi=\frac{\pi}{3}$ in the low- frequency range $10Hz\leq f\leq100Hz$}
\end{figure}
 The angular dependences of the total response function (\ref{eq: risposta totale 2 per magn})
of the Virgo and LIGO interferometers to the magnetic component of
the $\times$ polarization for $f=100Hz$ are shown in figure 9 and
10. %
\begin{figure}
\includegraphics{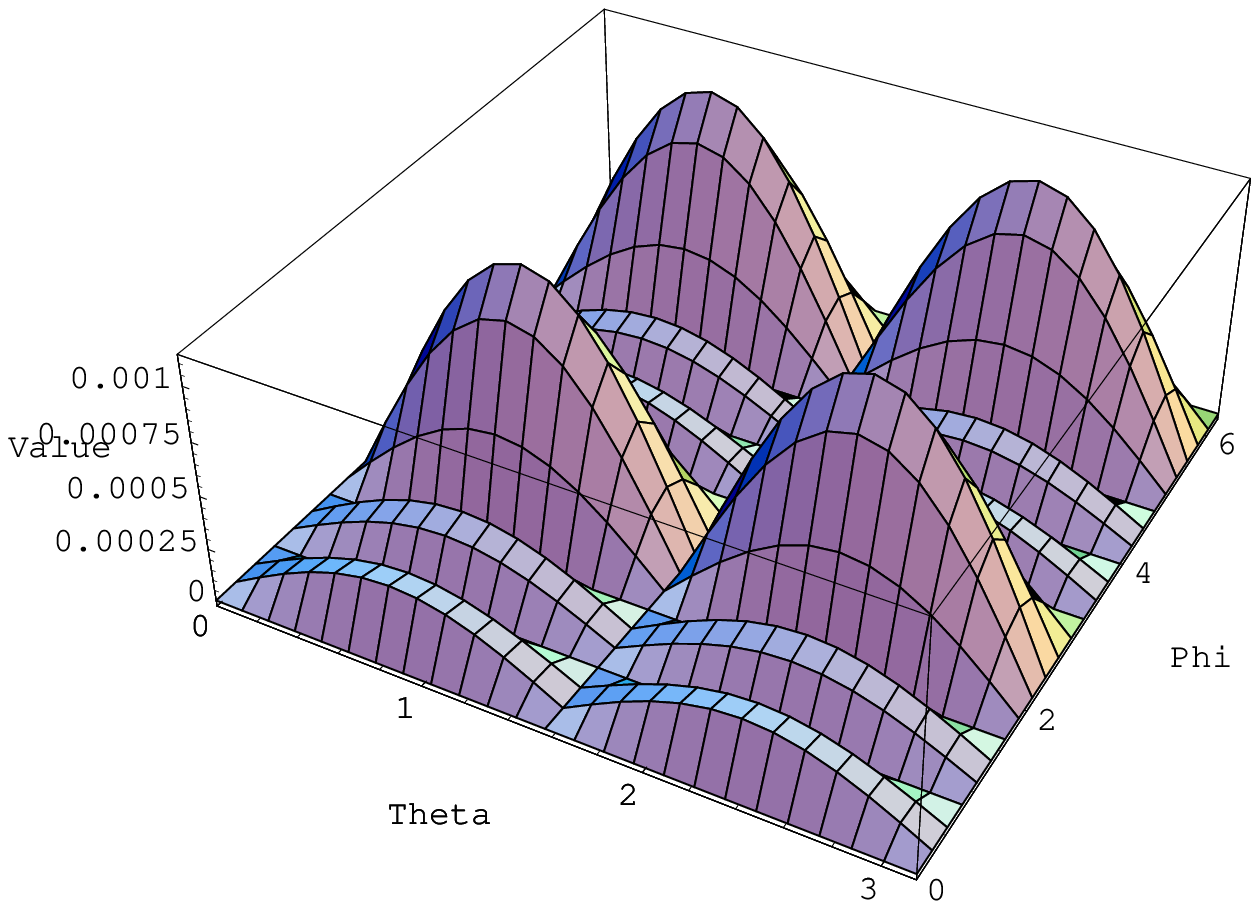}

\caption{the angular dependence of the total response function of the Virgo
interferometer to the magnetic component of the $\times$ polarization
for $f=100Hz$}
\end{figure}
\begin{figure}
\includegraphics{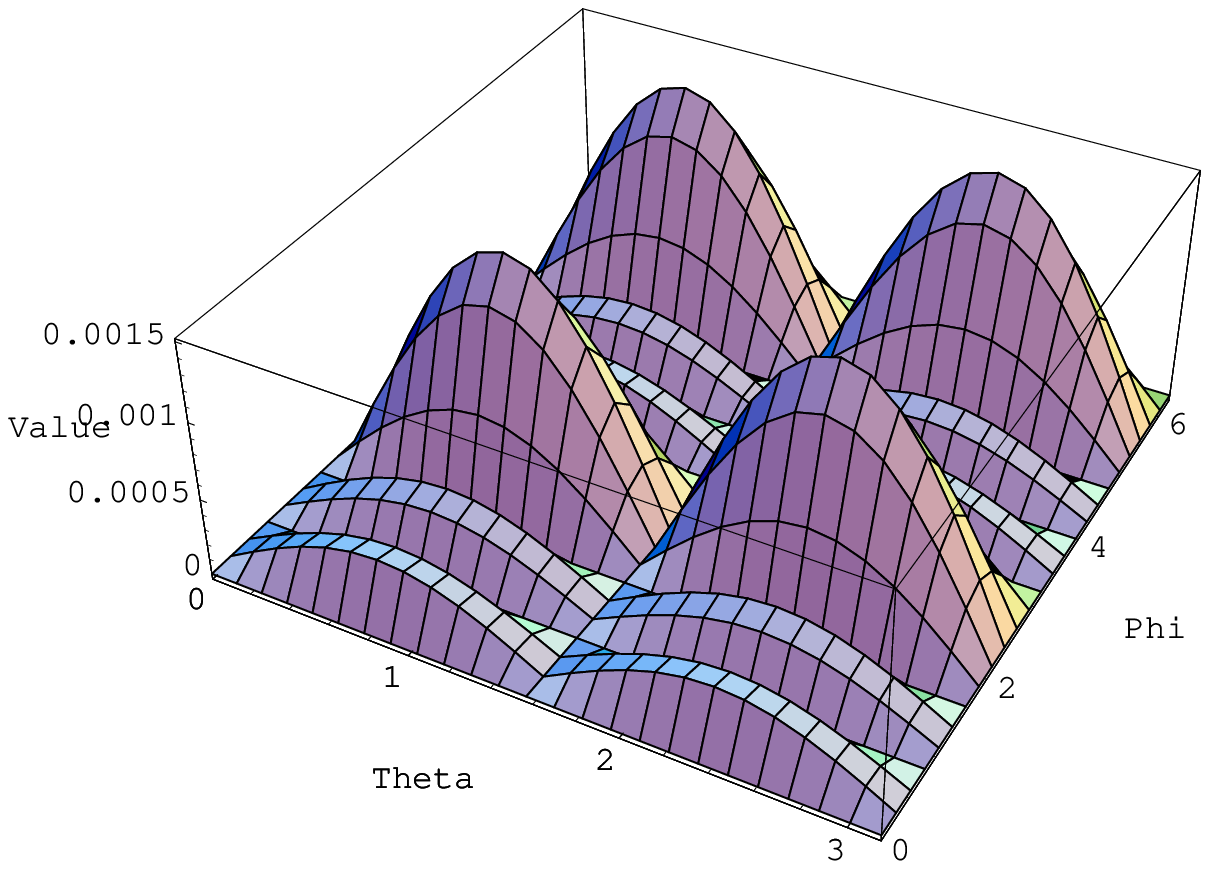}

\caption{the angular dependence of the total response function of the LIGO
interferometer to the magnetic component of the $\times$ polarization
for $f=100Hz$}
\end{figure}

\section{More accurate response functions for the magnetic components}

One can extend equations (\ref{eq: trasf. coord.}) in the form \cite{key-25,key-27}

\begin{equation}
\begin{array}{c}
t(t+z)=t_{tt}+\frac{1}{4}(x_{tt}^{2}-y_{tt}^{2})\dot{h}_{+}(t+z)-\frac{1}{2}x_{tt}y_{tt}\dot{h}_{\times}(t+z)\\
\\x(t+z)=x_{tt}+\frac{1}{2}x_{tt}h_{+}(t+z)-\frac{1}{2}y_{tt}h_{\times}(t+z)+\frac{1}{2}x_{tt}z_{tt}\dot{h}_{+}(t+z)-\frac{1}{2}y_{tt}z_{tt}\dot{h}_{\times}(t+z)\\
\\y(t+z)=y_{tt}+\frac{1}{2}y_{tt}h_{+}(t+z)-\frac{1}{2}x_{tt}h_{\times}(t+z)+\frac{1}{2}y_{tt}z_{tt}\dot{h}_{+}(t+z)-\frac{1}{2}x_{tt}z_{tt}\dot{h}_{\times}(t+z)\\
\\z(t+z)=z_{tt}-\frac{1}{4}(x_{tt}^{2}-y_{tt}^{2})\dot{h}_{+}(t+z)+\frac{1}{2}x_{tt}y_{tt}\dot{h}_{\times}(t+z).\end{array}\label{eq: trasf. coord. 2}\end{equation}

This is because for a large separation between the test masses (in
the case of Virgo the distance between the beam-splitter and the mirror
is three kilometers, four in the case of LIGO), one cannot compute
the coefficients of this transformation (components of the metric
and its first time derivative) along the central wordline of the local
observer, but a dependence from the position of the test masses is
needed \cite{key-25,key-27}. Thus, also equations (\ref{eq: Grishuk 0}),
(\ref{eq: traditional}), (\ref{eq: news}), (\ref{eq: Grishuk 01}),
(\ref{eq: Grishuk 1}) and (\ref{eq: du magn}) have to be modified
in the same way. In particular, we get \begin{equation}
u=L+\frac{1}{4}L^{2}A\dot{h}_{+}(t+u\sin\theta\cos\phi).\label{eq: du}\end{equation}

From eq. (\ref{eq: du}) we find that the displacements of the two
masses under the influence of the GW are

\begin{equation}
\delta u_{b}(t)=0\label{eq: spostamento beam-splitter acc}\end{equation}

and

\begin{equation}
\delta u_{m}(t)=\frac{1}{4}L^{2}A\dot{h}_{+}(t+L\sin\theta\cos\phi).\label{eq: spostamento mirror acc}\end{equation}

In this way, the relative displacement, which is defined by

\begin{equation}
\delta L(t)=\delta u_{m}(t)-\delta u_{b}(t)\label{eq: spostamento relativo acc}\end{equation}

gives

\begin{equation}
\frac{\delta T(t)}{T}=\frac{\delta L(t)}{L}=\frac{1}{4}LA\dot{h}_{+}(t+L\sin\theta\cos\phi).\label{eq: strain magnetico acc}\end{equation}
But we have the problem that, for the large separation between the
test masses, the definition (\ref{eq: spostamento relativo acc})
for relative displacements becomes unphysical because the two test
masses are taken at the same time and therefore cannot be in a casual
connection \cite{key-25,key-27}. We can write the correct definitions
using a the so called {}``bouncing photon method'': a photon can
be launched from the beam-splitter to be bounced back by the mirror
(Figure 1). This method has been generalized to scalar waves, angular
dependences and massive modes of GWs in \cite{key-2,key-9,key-25,key-26,key-27}. 

One obtains: 

\begin{equation}
\delta L_{1}(t)=\delta u_{m}(t)-\delta u_{b}(t-T_{1})\label{eq: corretto spostamento B.S. e M. acc}\end{equation}

and

\begin{equation}
\delta L_{2}(t)=\delta u_{m}(t-T_{2})-\delta u_{b}(t),\label{eq: corretto spostamento B.S. e M. 2 acc}\end{equation}

where $T_{1}$ and $T_{2}$ are the photon propagation times for the
forward and return trip correspondingly. According to the new definitions,
the displacement of one test mass is compared with the displacement
of the other at a later time to allow for finite delay from the light
propagation. We note that the propagation times $T_{1}$ and $T_{2}$
in eqs. (\ref{eq: corretto spostamento B.S. e M. acc}) and (\ref{eq: corretto spostamento B.S. e M. 2 acc})
can be replaced with the nominal value $T$ because the test mass
displacements are alredy first order in $h_{+}$ \cite{key-25,key-27}.
Thus, for the total change in the distance between the beam splitter
and the mirror in one round-trip of the photon, we get

\begin{equation}
\delta L_{r.t.}(t)=\delta L_{1}(t-T)+\delta L_{2}(t)=2\delta u_{m}(t-T)-\delta u_{b}(t)-\delta u_{b}(t-2T),\label{eq: variazione distanza propria acc}\end{equation}

and in terms of the amplitude of the GW:

\begin{equation}
\delta L_{r.t.}(t)=\frac{1}{2}L^{2}A\dot{h}_{+}(t+L\sin\theta\cos\phi-L).\label{eq: variazione distanza propria 2 acc}\end{equation}
The change in distance (\ref{eq: variazione distanza propria 2 acc})
leads to changes in the round-trip time for photons propagating between
the beam-splitter and the mirror:

\begin{equation}
\frac{\delta_{1}T(t)}{T}=\frac{1}{2}LA\dot{h}_{+}(t+L\sin\theta\cos\phi-L).\label{eq: variazione tempo proprio 1 acc}\end{equation}

\section{Effect of curved spacetime}

In the last calculation (variations in the photon round-trip time
which come from the motion of the test masses inducted by the magnetic
component of the $+$ polarization of the GW), we implicitly assumed
that the propagation of the photon between the beam-splitter and the
mirror of the interferometer is uniform as if it were moving in a
flat space-time. But the presence of the tidal forces indicates that
the space-time is curved. As a result, we have to consider one more
effect after the first discussed that requires spacial separation
\cite{key-25,key-27}. 

From equation (\ref{eq: spostamento mirror acc}) we get the tidal
acceleration of a test mass caused by the magnetic component of the
$+$ polarization of the GW in the $u$ direction \begin{equation}
\ddot{u}(t+u\sin\theta\cos\phi)=\frac{1}{4}L^{2}A\frac{\partial}{\partial t}\ddot{h}{}_{+}(t+u\sin\theta\cos\phi).\label{eq: acc}\end{equation}

Equivalently,  we can say that there is a gravitational potential
\cite{key-25,key-27}:

\begin{equation}
V(u,t)=-\frac{1}{4}L^{2}A\int_{0}^{u}\frac{\partial}{\partial t}\ddot{h}{}_{+}(t+l\sin\theta\cos\phi)dl,\label{eq:potenziale in gauge Lorentziana acc}\end{equation}

which generates the tidal forces, and that the motion of the test
mass is governed by the Newtonian equation

\begin{equation}
\ddot{\overrightarrow{r}}=-\bigtriangledown V.\label{eq: Newtoniana acc}\end{equation}

For the second effect, we consider the interval for photons propagating
along the $u$ -axis\begin{equation}
ds^{2}=g_{00}dt^{2}+du^{2}.\label{eq: metrica osservatore locale acc}\end{equation}

The condition for a null trajectory ($ds=0$) gives the coordinate
velocity of the photons 

\begin{equation}
v^{2}\equiv(\frac{du}{dt})^{2}=1+2V(t,u),\label{eq: velocita' fotone in gauge locale acc}\end{equation}

which to first order in $h_{+}$ is approximated by

\begin{equation}
v\approx\pm[1+V(t,u)],\label{eq: velocita fotone in gauge locale 2 acc}\end{equation}

with $+$ and $-$ for the forward and return trip respectively. If
we know the coordinate velocity of the photon, we can define the propagation
time for its travelling between the beam-splitter and the mirror:

\begin{equation}
T_{1}(t)=\int_{u_{b}(t-T_{1})}^{u_{m}(t)}\frac{du}{v}\label{eq:  tempo di propagazione andata gauge locale acc}\end{equation}

and

\begin{equation}
T_{2}(t)=\int_{u_{m}(t-T_{2})}^{u_{b}(t)}\frac{(-du)}{v}.\label{eq:  tempo di propagazione ritorno gauge locale acc}\end{equation}

The calculations of these integrals would be complicated because the
$u_{m}$ boundaries of them are changing with time:

\begin{equation}
u_{b}(t)=0\label{eq: variazione b.s. in gauge locale acc}\end{equation}

and

\begin{equation}
u_{m}(t)=L+\delta u_{m}(t).\label{eq: variazione specchio nin gauge locale acc}\end{equation}

But we note that, to first order in $h{}_{+}$, these contributions
can be approximated by $\delta L_{1}(t)$ and $\delta L_{2}(t)$ (see
eqs. (\ref{eq: corretto spostamento B.S. e M. acc}) and (\ref{eq: corretto spostamento B.S. e M. 2 acc})).
Thus, the combined effect of the varying boundaries is given by $\delta_{1}T(t)$
in eq. (\ref{eq: variazione tempo proprio 1 acc}). Then, we have
only to calculate the times for photon propagation between the fixed
boundaries: $0$ and $L$. We will denote such propagation times with
$\Delta T_{1,2}$ to distinguish from $T_{1,2}$. In the forward trip,
the propagation time between the fixed limits is

\begin{equation}
\Delta T_{1}(t)=\int_{0}^{L}\frac{du}{v(t',u)}\approx L-\int_{0}^{L}V(t',u)du,\label{eq:  tempo di propagazione andata  in gauge locale acc}\end{equation}

where $t'$ is the delay time (i.e. $t$ is the time at which the
photon arrives in the position $L$, so $L-u=t-t'$) which corresponds
to the unperturbed photon trajectory: 

\begin{center}$t'=t-(L-u)$. \end{center}

Similarly, the propagation time in the return trip is

\begin{equation}
\Delta T_{2}(t)=L-\int_{L}^{0}V(t',u)du,\label{eq:  tempo di propagazione andata  in gauge locale acc}\end{equation}

where now the delay time is given by

\begin{center}$t'=t-u$.\end{center}

The sum of $\Delta T_{1}(t-T)$ and $\Delta T_{2}(t)$ give us the
round-trip time for photons traveling between the fixed boundaries.
Then, we obtain the deviation of this round-trip time (distance) from
its unperturbed value $2T$:\begin{equation}
\begin{array}{c}
\delta_{2}T(t)=-\int_{0}^{L}[V(t-2L+u,u)du+\\
\\-\int_{L}^{0}V(t-u,u)]du,\end{array}\label{eq: variazione tempo proprio 2 acc}\end{equation}

and, using eq. (\ref{eq:potenziale in gauge Lorentziana acc}), it
is

\begin{equation}
\begin{array}{c}
\delta_{2}T(t)=\frac{1}{4}L^{2}A\int_{0}^{L}[\int_{0}^{u}\frac{\partial}{\partial t}\ddot{h}_{+}(t-2T+l(1+\sin\theta\cos\phi))dl+\\
\\-\int_{0}^{u}\frac{\partial}{\partial t}\ddot{h}_{+}(t-l(1-\sin\theta\cos\phi)dl]du.\end{array}\label{eq: variazione tempo proprio 2 rispetto h acc}\end{equation}

Thus, the total round-trip proper distance in presence of the magnetic
component of the $+$ polarization of the GW is:

\begin{equation}
T_{t}=2T+\delta_{1}T+\delta_{2}T,\label{eq: round-trip  totale in gauge locale acc}\end{equation}

and\begin{equation}
\delta T_{u}=T_{t}-2T=\delta_{1}T+\delta_{2}T\label{eq:variaz round-trip totale in gauge locale acc}\end{equation}

is the total variation of the proper time (distance) for the round-trip
of the photon in presence of the magnetic component of the GW in the
$u$ direction.

Using eqs. (\ref{eq: variazione tempo proprio 1 acc}), (\ref{eq: variazione tempo proprio 2 rispetto h acc})
and the Fourier transform of $h_{+}$ defined by (\ref{eq: trasformata di fourier magn}),
the quantity (\ref{eq:variaz round-trip totale in gauge locale acc})
can be computed in the frequency domain: 

\begin{equation}
\tilde{\delta}T_{u}(\omega)=\tilde{\delta}_{1}T(\omega)+\tilde{\delta}_{2}T(\omega)\label{eq:variaz round-trip totale in gauge locale 2 acc}\end{equation}

where

\begin{equation}
\tilde{\delta}_{1}T(\omega)=-i\omega\exp[i\omega L(1-\sin\theta\cos\phi)]\frac{L^{2}A}{2}\tilde{h}_{+}(\omega)\label{eq: dt 1 omega acc}\end{equation}

\begin{equation}
\begin{array}{c}
\tilde{\delta}_{2}T(\omega)=\frac{i\omega L^{2}A}{4}[\frac{-1+\exp[i\omega L(1-\sin\theta\cos\phi)]-iL\omega(1-\sin\theta\cos\phi)}{(1-\sin\theta\cos\phi)^{2}}+\\
\\+\frac{\exp(2i\omega L)(1-\exp[i\omega L(-1-\sin\theta\cos\phi)]-iL\omega(1+\sin\theta\cos\phi)}{(-1-\sin\theta\cos\phi)^{2}}]\tilde{h}_{+}(\omega).\end{array}\label{eq: dt 2 omega acc}\end{equation}

In the above computation the derivation and translation theorems of
the Fourier transform have been used. In this way, the response function
of the $u$ arm of the interferometer to the magnetic component of
the $+$ polarization of the GW is

\begin{equation}
\begin{array}{c}
H_{u}^{+}(\omega)\equiv\frac{\tilde{\delta}T_{u}(\omega)}{L\tilde{h}_{+}(\omega)}=\\
\\=-i\omega\exp[i\omega L(1-\sin\theta\cos\phi)]\frac{LA}{2}+\\
\\\frac{i\omega LA}{4}[\frac{-1+\exp[i\omega L(1-\sin\theta\cos\phi)]-iL\omega(1-\sin\theta\cos\phi)}{(1-\sin\theta\cos\phi)^{2}}+\\
\\+\frac{\exp(2i\omega L)(1-\exp[i\omega L(-1-\sin\theta\cos\phi)]-iL\omega(1+\sin\theta\cos\phi)}{(-1-\sin\theta\cos\phi)^{2}}].\end{array}\label{eq: risposta u acc}\end{equation}

\section{Computation for the $v$ arm}

The computation for the $v$ arm is parallel to the one above. Using
eqs. (\ref{eq: Grishuk 1}), (\ref{eq: rotazione magn}) and (\ref{eq: rotazione 2 magn})
the coordinate of the mirror in the $v$ arm is:

\begin{equation}
v=L+\frac{1}{4}L^{2}B\dot{h}_{+}(t+v\sin\theta\sin\phi).\label{eq: dv acc}\end{equation}

Thus, with the same way of thinking of previous Sections, we get variations
in the photon round-trip time which come from the motion of the beam-splitter
and the mirror in the $v$ direction:

\begin{equation}
\frac{\delta_{1}T(t)}{T}=\frac{1}{2}LB\dot{h}_{+}(t+L\sin\theta\sin\phi-L),\label{eq: variazione tempo proprio 1 in v acc}\end{equation}

while the second contribute (propagation in a curve spacetime) will
be 

\begin{equation}
\begin{array}{c}
\delta_{2}T(t)=\frac{1}{4}L^{2}B\int_{0}^{L}[\int_{0}^{u}\frac{\partial}{\partial t}\ddot{h}_{+}(t-2T+l(1-\sin\theta\sin\phi))dl+\\
\\-\int_{0}^{u}\frac{\partial}{\partial t}\ddot{h}_{+}(t-l(1-\sin\theta\sin\phi)dl]du,\end{array}\label{eq: variazione tempo proprio 2 rispetto h in v acc}\end{equation}

and the total response function of the $v$ arm for the magnetic component
of the $+$ polarization of GWs is given by\begin{equation}
\begin{array}{c}
H_{v}^{+}(\omega)\equiv\frac{\tilde{\delta}T_{u}(\omega)}{L\tilde{h}_{+}(\omega)}=\\
\\=-i\omega\exp[i\omega L(1-\sin\theta\sin\phi)]\frac{LB}{2}+\\
\\+\frac{i\omega LB}{4}[\frac{-1+\exp[i\omega L(1-\sin\theta\sin\phi)]-iL\omega(1-\sin\theta\sin\phi)}{(1-\sin\theta\cos\phi)^{2}}+\\
\\+\frac{\exp(2i\omega L)(1-\exp[i\omega L(-1-\sin\theta\sin\phi)]-iL\omega(1+\sin\theta\sin\phi)}{(-1-\sin\theta\sin\phi)^{2}}].\end{array}\label{eq: risposta v acc}\end{equation}

\section{The total response function of an interferometer for the $+$ polarization}

The total response function for the magnetic component of the $+$
polarization is given by the difference of the two response function
of the two arms:\begin{equation}
H_{tot}^{+}(\omega)\equiv H_{u}^{+}(\omega)-H_{v}^{+}(\omega),\label{eq: risposta totale acc}\end{equation}

and, using eqs. (\ref{eq: risposta u acc}) and (\ref{eq: risposta v acc}),
we obtain a complicated formula\begin{equation}
\begin{array}{c}
H_{tot}^{+}(\omega)=\frac{\tilde{\delta}T_{tot}(\omega)}{L\tilde{h}_{+}(\omega)}=\\
\\=-i\omega\exp[i\omega L(1-\sin\theta\cos\phi)]\frac{LA}{2}+\frac{LB}{2}i\omega\exp[i\omega L(1-\sin\theta\sin\phi)]\\
\\-\frac{i\omega LA}{4}[\frac{-1+\exp[i\omega L(1-\sin\theta\cos\phi)]-iL\omega(1-\sin\theta\cos\phi)}{(1-\sin\theta\cos\phi)^{2}}\\
\\+\frac{\exp(2i\omega L)(1-\exp[i\omega L(-1-\sin\theta\cos\phi)]-iL\omega(1+\sin\theta\cos\phi)}{(-1-\sin\theta\cos\phi)^{2}}]+\\
\\+\frac{i\omega LB}{4}[\frac{-1+\exp[i\omega L(1-\sin\theta\sin\phi)]-iL\omega(1-\sin\theta\sin\phi)}{(1-\sin\theta\cos\phi)^{2}}+\\
\\+\frac{\exp(2i\omega L)(1-\exp[i\omega L(-1-\sin\theta\sin\phi)]-iL\omega(1+\sin\theta\sin\phi)}{(-1-\sin\theta\sin\phi)^{2}}],\end{array}\label{eq: risposta totale 2 acc}\end{equation}

that, in the low freuencies limit is in perfect agreement with the
result of Baskaran and Grishchuk (eq. 49 of \cite{key-24}), i.e.
with eq. (\ref{eq: risposta totale magn}): \begin{equation}
H_{tot}^{+}(\omega\rightarrow0)=\frac{1}{4}\sin\theta[(\cos^{2}\theta+\sin2\phi\frac{1+\cos^{2}\theta}{2})](\cos\phi-\sin\phi).\label{eq: risposta totale bassa acc}\end{equation}

In figures 11 and 12 the angular dependences of the total response
function (\ref{eq: risposta totale 2 acc}) of the Virgo and LIGO
interferometers to the magnetic component of the $+$ polarization
of GWs at the frequency $f=8000Hz$ are respectively shown. This frequency
falls in the high-frequency portion of the interferometers sensitivity
band, thus, the {}``magnetic'' contribution becomes quit important.
In fact, figures 11 and 12 show that it can go over the 10\% of the
total signal.%
\begin{figure}
\begin{center}\includegraphics{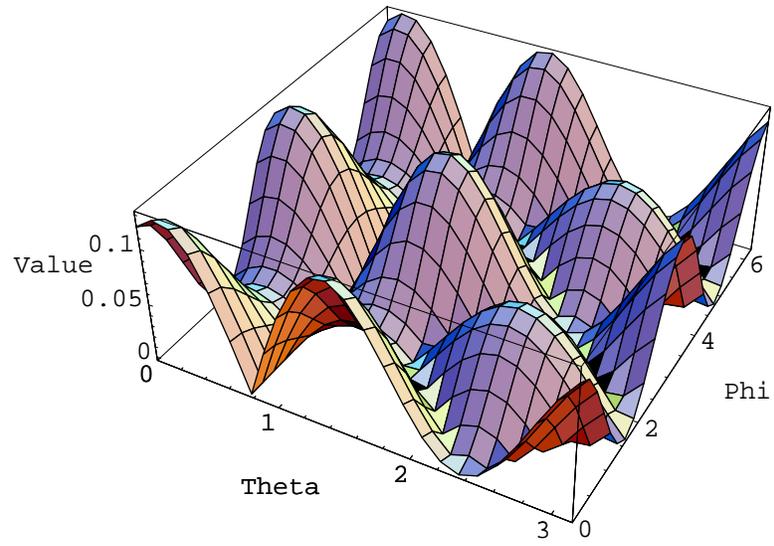}\end{center}

\caption{the angular dependence of the response function of the Virgo interferometer
to the magnetic component of the $+$ polarization for $f=8000Hz$}
\end{figure}
\begin{figure}
\begin{center}\includegraphics{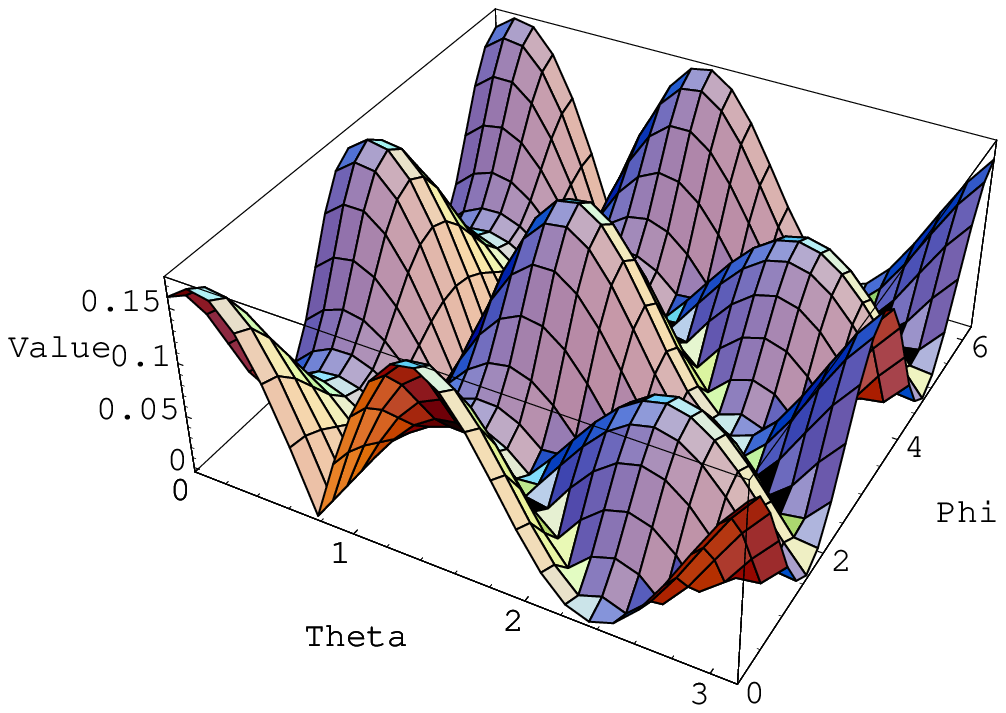}\end{center}

\caption{the angular dependence of the response function of the LIGO interferometer
to the magnetic component of the $+$ polarization for $f=8000Hz$}
\end{figure}

\section{Analysis for the $\times$ polarization}

The analysis can be generalized for the magnetic component of the
$\times$ polarization too. In this case, using equations (\ref{eq: trasf. coord. 2}),
(\ref{eq: rotazione magn}) and (\ref{eq: rotazione 2 magn}) the
$u$ coordinate of the mirror situated in the $u$ arm of the interferometer
is given by \begin{equation}
u=L+\frac{1}{4}L^{2}C\dot{h}_{\times}(t+u\sin\theta\cos\phi).\label{eq: du C}\end{equation}

while the $v$ coordinate of the mirror situated in the $v$ arm of
the interferometer is given by \begin{equation}
v=L+\frac{1}{4}L^{2}D\dot{h}_{\times}(t+v\sin\theta\sin\phi).\label{eq: dv  D}\end{equation}

Thus, with an analysis similar to the one of previous Sections, it
is possible to show that the total response function of the interferometer
for the magnetic component of the $\times$ polarization of GWs is\begin{equation}
\begin{array}{c}
H_{tot}^{\times}(\omega)=\frac{\tilde{\delta}T_{tot}(\omega)}{L\tilde{h}_{\times}(\omega)}=\\
\\=-i\omega\exp[i\omega L(1-\sin\theta\cos\phi)]\frac{LC}{2}+\frac{LD}{2}i\omega\exp[i\omega L(1-\sin\theta\sin\phi)]\\
\\-\frac{i\omega LC}{4}[\frac{-1+\exp[i\omega L(1-\sin\theta\cos\phi)]-iL\omega(1-\sin\theta\cos\phi)}{(1-\sin\theta\cos\phi)^{2}}\\
\\+\frac{\exp(2i\omega L)(1-\exp[i\omega L(-1-\sin\theta\cos\phi)]-iL\omega(1+\sin\theta\cos\phi)}{(-1-\sin\theta\cos\phi)^{2}}]+\\
\\+\frac{i\omega LD}{4}[\frac{-1+\exp[i\omega L(1-\sin\theta\sin\phi)]-iL\omega(1-\sin\theta\sin\phi)}{(1-\sin\theta\cos\phi)^{2}}+\\
\\+\frac{\exp(2i\omega L)(1-\exp[i\omega L(-1-\sin\theta\sin\phi)]-iL\omega(1+\sin\theta\sin\phi)}{(-1-\sin\theta\sin\phi)^{2}}],\end{array}\label{eq: risposta totale 2 per acc}\end{equation}

that, in the low frequencies limit, is in perfect agreement with the
result of Baskaran and Grishchuk (eq. 50 of \cite{key-24}) and with
eq. (\ref{eq: risposta totale 2 per magn}): \begin{equation}
H_{tot}^{\times}(\omega\rightarrow0)=\frac{1}{4}\sin2\phi(\cos\phi+\sin\phi)\cos\theta.\label{eq: risposta totale bassa per acc}\end{equation}
In figure 13 and 14 the angular dependences of the total response
function (\ref{eq: risposta totale 2 per acc}) of the Virgo and LIGO
interferometers to the magnetic component of the $\times$ polarization
of GWs at the frequency $f=8000Hz$ are respectively shown. 

\begin{figure}
\begin{center}\includegraphics{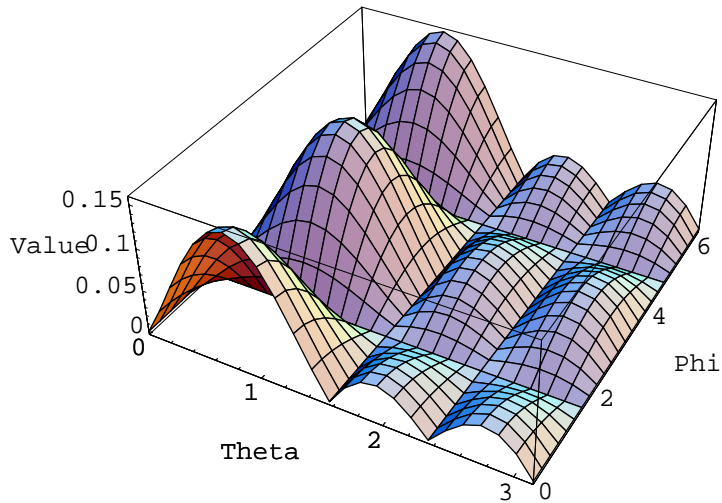}\end{center}

\caption{the angular dependence of the response function of the Virgo interferometer
to the magnetic component of the $\times$ polarization for $f=8000Hz$}
\end{figure}
\begin{figure}
\begin{center}\includegraphics{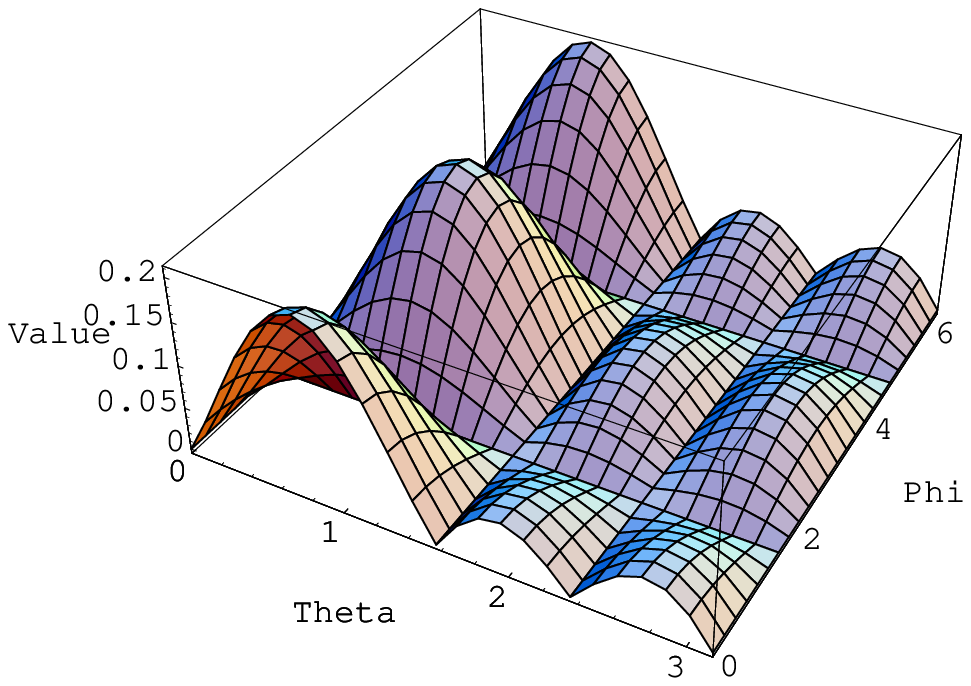}\end{center}

\caption{the angular dependence of the response function of the LIGO interferometer
to the magnetic component of the $\times$ polarization for $f=8000Hz$}
\end{figure}
The figures show the importance of the {}``magnetic'' contribution
in the high-frequency portion of the interferometers sensitivity band
in this case ($\times$ polarization) too.

Because the response functions to the {}``magnetic'' components
grow with frequency, as it is shown in eqs. (\ref{eq: risposta totale 2 acc})
and (\ref{eq: risposta totale 2 per acc}), one could think that the
part of signal which arises from the magnetic components could in
principle become the dominant part of the signal at high frequencies
(see also \cite{key-25,key-27}), and, in principle, extend the frequency
range of interferometers. But, to understand if this is correct, one
has to use the full theory of gravitational waves.

\section{The total response function of interferometers in the full theory
of gravitational waves}

The low-frequencies approximation, used in Sections 3 and 4 to show
that the {}``magnetic'' and {}``electric'' contributions to the
response functions can be identified without ambiguity in the longh-wavelengths
regime \cite{key-26,key-27,key-24}, is sufficient only for ground
based interferometers, for which the condition $f\ll1/L$ is in general
satisfied. For space-based interferometers, for which the above condition
is not satisfied in the high-frequency portion of the sensitivity
band \cite{key-24,key-25,key-26,key-27}, the response functions of
Sections 8-9 give a better approximation \cite{key-24,key-25,key-26,key-27}.
But, to compute the correct total response function, without any approximation
in distance and/or frequency, the full theory of gravitational waves
has to be used \cite{key-2,key-26}.

In this Section, the variation of the proper distance that a photon
covers to make  a round-trip from the beam-splitter to the mirror
of an interferometer is computed with the gauge choice (\ref{eq: metrica TT totale})
(see also \cite{key-2,key-26}). In this case, one does not need the
coordinate transformation (\ref{eq: trasf. coord.}) from the TT coordinates
to the frame of the local observer. Thus, with a treatment similar
to the one of \cite{key-2,key-26}, the analysis is translated in
the frequency domain and the general response functions are obtained.

A special property of the TT gauge is that an inertial test mass initially
at rest in these coordinates, remains at rest throughout the entire
passage of the GW \cite{key-2,key-26}. Here we have to clarify the
use of words {}`` at rest'': we want to mean that the coordinates
of the test mass do not change in the presence of the GW. The proper
distance between the beam-splitter and the mirror of the interferometer
changes even though their coordinates remain the same \cite{key-2,key-26}.

We start from the $+$ polarization. Labelling the coordinates of
the TT gauge with $t,x,y,z$ for a sake of simplicity, the line element
(\ref{eq: metrica TT totale}) becomes:

\begin{equation}
ds^{2}=-dt^{2}+dz^{2}+[1+h_{+}(t+z)]dx^{2}+[1+h_{+}(t+z)]dy^{2}.\label{eq: metrica polarizzazione + full}\end{equation}

But the arms of the interferometer are in the $\overrightarrow{u}$
and $\overrightarrow{v}$ directions, while the $x,y,z$ frame is
the proper frame of the propagating GW. 

The coordinate transformation for the metric tensor is \cite{key-2,key-26}:

\begin{equation}
g^{ik}=\frac{\partial x^{i}}{\partial x'^{l}}\frac{\partial x^{k}}{\partial x'^{m}}g'^{lm}.\label{eq: trasformazione metrica full}\end{equation}

By using eq. (\ref{eq: rotazione magn}), (\ref{eq: rotazione 2 magn})
and (\ref{eq: trasformazione metrica full}), in the new rotated frame
the line element (\ref{eq: metrica polarizzazione + full}) in the
$\overrightarrow{u}$ direction becomes:

\begin{equation}
ds^{2}=-dt^{2}+[1+(\cos^{2}\theta\cos^{2}\phi-\sin^{2}\phi)h_{+}(t+u\sin\theta\cos\phi)]du^{2}.\label{eq: metrica + lungo u full}\end{equation}

The condition for null geodesics ($ds^{2}=0$) in eq. (\ref{eq: metrica + lungo u full})
gives the coordinate velocity of the photon:

\begin{equation}
v^{2}\equiv(\frac{du}{dt})^{2}=\frac{1}{[1+(\cos^{2}\theta\cos^{2}\phi-\sin^{2}\phi)h_{+}(t+u\sin\theta\cos\phi)]}.\label{eq: velocita' fotone u full}\end{equation}

We recall that the beam splitter is located in the origin of the new
coordinate system (i.e. $u_{b}=0$, $v_{b}=0$, $w_{b}=0$). The coordinates
of the beam-splitter $u_{b}=0$ and of the mirror $u_{m}=L$ do not
change under the influence of the GW, thus, the duration of the forward
trip can be written as

\begin{equation}
T_{1}(t)=\int_{0}^{L}\frac{du}{v(t'+u\sin\theta\cos\phi)},\label{eq: durata volo full}\end{equation}

with 

\begin{center}$t'=t-(L-u)$.\end{center}

In the last equation $t'$ is the delay time (see Section 6).

At first order in $h_{+}$ this integral can be approximated with

\begin{equation}
T_{1}(t)=T+\frac{\cos^{2}\theta\cos^{2}\phi-\sin^{2}\phi}{2}\int_{0}^{L}h_{+}(t'+u\sin\theta\cos\phi)du,\label{eq: durata volo andata approssimata u full}\end{equation}

where

\begin{center}$T=L$ \end{center}

is the transit time of the photon in absence of the GW. Similiarly,
the duration of the return trip will be\begin{equation}
T_{2}(t)=T+\frac{\cos^{2}\theta\cos^{2}\phi-\sin^{2}\phi}{2}\int_{L}^{0}h_{+}(t'+u\sin\theta\cos\phi)(-du),\label{eq: durata volo ritorno approssimata u full}\end{equation}

though now the delay time is 

\begin{center}$t'=t-(u-l)$.\end{center}

The round-trip time will be the sum of $T_{2}(t)$ and $T_{1}[t-T_{2}(t)]$.
The latter can be approximated by $T_{1}(t-T)$ because the difference
between the exact and the approximate values is second order in $h_{+}$.
Then, to first order in $h_{+}$, the duration of the round-trip will
be

\begin{equation}
T_{r.t.}(t)=T_{1}(t-T)+T_{2}(t).\label{eq: durata round trip full}\end{equation}

By using eqs. (\ref{eq: durata volo andata approssimata u full})
and (\ref{eq: durata volo ritorno approssimata u full}) one sees
immediately that deviations of this round-trip time (i.e. proper distance)
from its unperturbed value are given by

\begin{equation}
\begin{array}{c}
\delta T(t)=\frac{\cos^{2}\theta\cos^{2}\phi-\sin^{2}\phi}{2}\int_{0}^{L}[h_{+}(t-2T-u(1-\sin\theta\cos\phi))+\\
\\+h_{+}(t+u(1+\sin\theta\cos\phi))]du.\end{array}\label{eq: variazione temporale in u full}\end{equation}

Now, using the Fourier transform of the $+$ polarization of the field,
defined by eq. (\ref{eq: trasformata di fourier magn}), one obtains
in the frequency domain:

\begin{equation}
\delta\tilde{T}(\omega)=\frac{1}{2}(\cos^{2}\theta\cos^{2}\phi-\sin^{2}\phi)\tilde{H}_{u}(\omega,\theta,\phi)\tilde{h}_{+}(\omega),\label{eq: segnale in frequenza lungo u full}\end{equation}

where

\begin{equation}
\begin{array}{c}
\tilde{H}_{u}(\omega,\theta,\phi)=\frac{-1+\exp(2i\omega L)}{2i\omega(1+\sin^{2}\theta\cos^{2}\phi)}+\\
\\+\frac{-\sin\theta\cos\phi((1+\exp(2i\omega L)-2\exp i\omega L(1-\sin\theta\cos\phi)))}{2i\omega(1+\sin\theta\cos^{2}\phi)}\end{array}\label{eq: fefinizione Hu full}\end{equation}

and we immediately see that $\tilde{H}_{u}(\omega,\theta,\phi)\rightarrow L$
when $\omega\rightarrow0$.

Thus, the total response function of the $u$ arm of the interferometer
to the $+$ component is:

\begin{equation}
\Upsilon_{u}^{+}(\omega)=\frac{(\cos^{2}\theta\cos^{2}\phi-\sin^{2}\phi)}{2L}\tilde{H}_{u}(\omega,\theta,\phi),\label{eq: risposta + lungo u full}\end{equation}

where $2L=2T$ is the round-trip time in absence of gravitational
waves.

In the same way, the line element (\ref{eq: metrica polarizzazione + full})
in the $\overrightarrow{v}$ direction becomes:

\begin{equation}
ds^{2}=-dt^{2}+[1+(\cos^{2}\theta\sin^{2}\phi-\cos^{2}\phi)h_{+}(t+v\sin\theta\sin\phi)]dv^{2},\label{eq: metrica + lungo v full}\end{equation}

and the response function of the $v$ arm of the interferometer to
the $+$ polarization is: 

\begin{equation}
\Upsilon_{v}^{+}(\omega)=\frac{(\cos^{2}\theta\sin^{2}\phi-\cos^{2}\phi)}{2L}\tilde{H}_{v}(\omega,\theta,\phi)\label{eq: risposta + lungo v full}\end{equation}

where, now 

\begin{equation}
\begin{array}{c}
\tilde{H}_{v}(\omega,\theta,\phi)=\frac{-1+\exp(2i\omega L)}{2i\omega(1+\sin^{2}\theta\sin^{2}\phi)}+\\
\\+\frac{-\sin\theta\sin\phi((1+\exp(2i\omega L)-2\exp i\omega L(1-\sin\theta\sin\phi)))}{2i\omega(1+\sin^{2}\theta\sin^{2}\phi)},\end{array}\label{eq: fefinizione Hv full}\end{equation}

with $\tilde{H}_{v}(\omega,\theta,\phi)\rightarrow L$ when $\omega\rightarrow0$.
In this case the variation of the distance (time) is\begin{equation}
\delta\tilde{T}(\omega)=\frac{1}{2}(\cos^{2}\theta\cos^{2}\phi-\cos^{2}\phi)\tilde{H}_{v}(\omega,\theta,\phi)\tilde{h}_{+}(\omega).\label{eq: segnale in frequenza lungo v full}\end{equation}

From equations (\ref{eq: segnale in frequenza lungo u full}) and
(\ref{eq: segnale in frequenza lungo v full}), the total lengths
of the two arms in presence of the $+$ polarization of the GW and
in the frequency domain are: \begin{equation}
\tilde{T}_{u}(\omega)=\frac{1}{2}(\cos^{2}\theta\cos^{2}\phi-\sin^{2}\phi)\tilde{H}_{u}(\omega,\theta,\phi)\tilde{h}_{+}(\omega)+T\label{eq: lunghezza u full}\end{equation}

and\begin{equation}
\tilde{T}_{v}(\omega)=\frac{1}{2}(\cos^{2}\theta\cos^{2}\phi-\cos^{2}\phi)\tilde{H}_{v}(\omega,\theta,\phi)\tilde{h}_{+}(\omega)+T,\label{eq: lunghezza v full}\end{equation}

that are particular cases of the more general equation (39) in \cite{key-24}.

Thus, the total frequency-dependent response function (i.e. the detector
pattern) of an interferometer to the $+$ polarization of the GW is:

\begin{equation}
\begin{array}{c}
\tilde{H}^{+}(\omega)=\Upsilon_{u}^{+}(\omega)-\Upsilon_{v}^{+}(\omega)=\\
\\=\frac{(\cos^{2}\theta\cos^{2}\phi-\sin^{2}\phi)}{2L}\tilde{H}_{u}(\omega,\theta,\phi)+\\
\\-\frac{(\cos^{2}\theta\sin^{2}\phi-\cos^{2}\phi)}{2L}\tilde{H}_{v}(\omega,\theta,\phi)\end{array}\label{eq: risposta totale Virgo + full}\end{equation}

that, in the low frequencies limit ($\omega\rightarrow0$) is in perfect
agreement with the detector pattern of eq. (46) in \cite{key-24},
if one retains the first two terms of the expansion:

\begin{equation}
\begin{array}{c}
\tilde{H}^{+}(\omega\rightarrow0)=\frac{1}{2}(1+\cos^{2}\theta)\cos2\phi+\\
\\-\frac{1}{4}i\omega L\sin\theta[(\cos^{2}\theta+\sin2\phi\frac{1+\cos^{2}\theta}{2})](\cos\phi-\sin\phi).\end{array}\label{eq: risposta totale approssimata full}\end{equation}

This result also confirms that the magnetic contribution to the variation
of the distance is an universal phenomenon because it has been obtained
starting from the full theory of gravitational waves in the TT gauge
(see also \cite{key-2,key-26,key-27}).

The same analysis can be now performed for the $\times$ polarization.
In this case, from eq. (\ref{eq: metrica TT totale}) the line element
is:

\begin{equation}
ds^{2}=-dt^{2}+dz^{2}+dx^{2}+dy^{2}+2h_{\times}(t+z)dxdy,\label{eq: metrica polarizzazione per full}\end{equation}

and, by using eqs. (\ref{eq: rotazione magn}), (\ref{eq: rotazione 2 magn})
and (\ref{eq: trasformazione metrica full}), the line element (\ref{eq: metrica polarizzazione per full})
in the $u$ direction and in the new rotated frame becomes:

\begin{equation}
ds^{2}=-dt^{2}+[1-2\cos\theta\cos\phi\sin\phi h_{\times}(t+u\sin\theta\cos\phi)]du^{2}.\label{eq: metrica per  lungo u full}\end{equation}

Then, the response function of the $u$ arm of the interferometer
to the $\times$ polarization is:

\begin{equation}
\Upsilon_{u}^{\times}(\omega)=\frac{-\cos\theta\cos\phi\sin\phi}{L}\tilde{H}_{u}(\omega,\theta,\phi),\label{eq: risposta per lungo u full}\end{equation}

while the line element (\ref{eq: metrica polarizzazione per full})
in the $v$ direction becomes:

\begin{equation}
ds^{2}=-dt^{2}+[1+2\cos\theta\cos\phi\sin\phi h_{\times}(t+u\sin\theta\sin\phi)]dv^{2}\label{eq: metrica per  lungo v full}\end{equation}

and the response function of the $v$ arm of the interferometer to
the $\times$ polarization is:

\begin{equation}
\Upsilon_{v}^{\times}(\omega)=\frac{\cos\theta\cos\phi\sin\phi}{L}\tilde{H}_{v}(\omega,\theta,\phi).\label{eq: risposta per lungo v full}\end{equation}

Thus, the total frequency-dependent response function of an interferometer
to the $\times$ polarization is:

\begin{equation}
\tilde{H}^{\times}(\omega)=\frac{-\cos\theta\cos\phi\sin\phi}{L}[\tilde{H}_{u}(\omega,\theta,\phi)+\tilde{H}_{v}(\omega,\theta,\phi)],\label{eq: risposta totale Virgo per full}\end{equation}

that, in the low frequencies limit ($\omega\rightarrow0$), is in
perfect agreement with the detector pattern of eq. (46) of \cite{key-24},
if one retains the first two terms of the expansion:

\begin{equation}
\tilde{H}^{\times}(\omega\rightarrow0)=-\cos\theta\sin2\phi-i\omega L\sin2\phi(\cos\phi+\sin\phi)\cos\theta.\label{eq: risposta totale approssimata 2 full}\end{equation}

The total lengths of the two arms in presence of the $\times$ polarization
and in the frequency domain are: \begin{equation}
\tilde{T}_{u}(\omega)=(\cos\theta\cos\phi\sin\phi)\tilde{H}_{u}(\omega,\theta,\phi)\tilde{h}_{\times}(\omega)+T\label{eq: lunghezza u full}\end{equation}

and\begin{equation}
\tilde{T}_{v}(\omega)=(-\cos\theta\cos\phi\sin\phi)\tilde{H}_{v}(\omega,\theta,\phi)\tilde{h}_{\times}(\omega)+T,\label{eq: lunghezza v full}\end{equation}

that also are particular cases of the more general equation (39) of
\cite{key-24}. The total low frequencies response functions of eqs.
(\ref{eq: risposta totale approssimata full}) and (\ref{eq: risposta totale approssimata 2 full})
are more accurate than the {}``traditional'' ones of \cite{key-29,key-30,key-31},
because our equations include the {}``magnetic'' contribution.

Thus, the obtained results confirm the presence and importance of
the so-called {}``magnetic'' components of GWs and the fact that
they have to be taken into account in the context of the total response
functions of interferometers for GWs propagating from arbitrary directions. 

The importance of the presented results is due to the fact that in
this case the limit where the wavelenght is shorter than the lenght
between the splitter mirror and test masses is calculated. The signal
drops off the regime, while the calculation agrees with previous calculations
for longer wavelenghts \cite{key-24,key-25}. The contribution is
important expecially in the high-frequency portion of the sensitivity
band.

In fact, one can see the pronounced difference between the {}``traditional''
low-frequency approximation angular pattern of the Virgo interferometer
for the $+$ polarization, i.e $\frac{1}{2}(1+\cos^{2}\theta)\cos2\phi$
as it is computed in \cite{key-29,key-30,key-31}, which is shown
in Figure 15, and the frequency-dependent angular pattern (\ref{eq: risposta totale Virgo + full}),
which is shown in Figure 16 at a frequency of 8000 Hz, i.e. a frequency
which falls \textit{in the high-frequency portion of the sensitivity
band.} The same angular patterns are shown in Figures 17 and 18 for
the LIGO interferometer. The difference between the low-frequency
approximation angular patterns and the frequency-dependent ones is
important for the $\times$ polarization too, as it is shown in Figures
19, 20 for Virgo and in Figures 21, 22 for LIGO.

\begin{figure}
\includegraphics{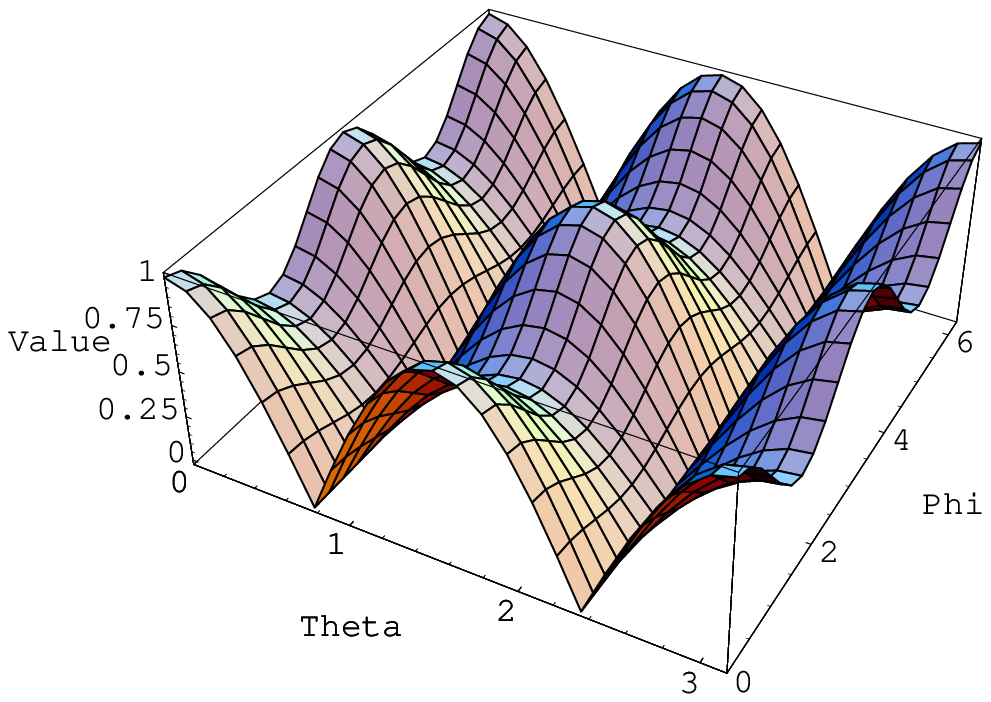}

\caption{The low-frequency angular dependence to the $+$ polarization for
the Virgo interferometer}
\end{figure}
\begin{figure}
\includegraphics{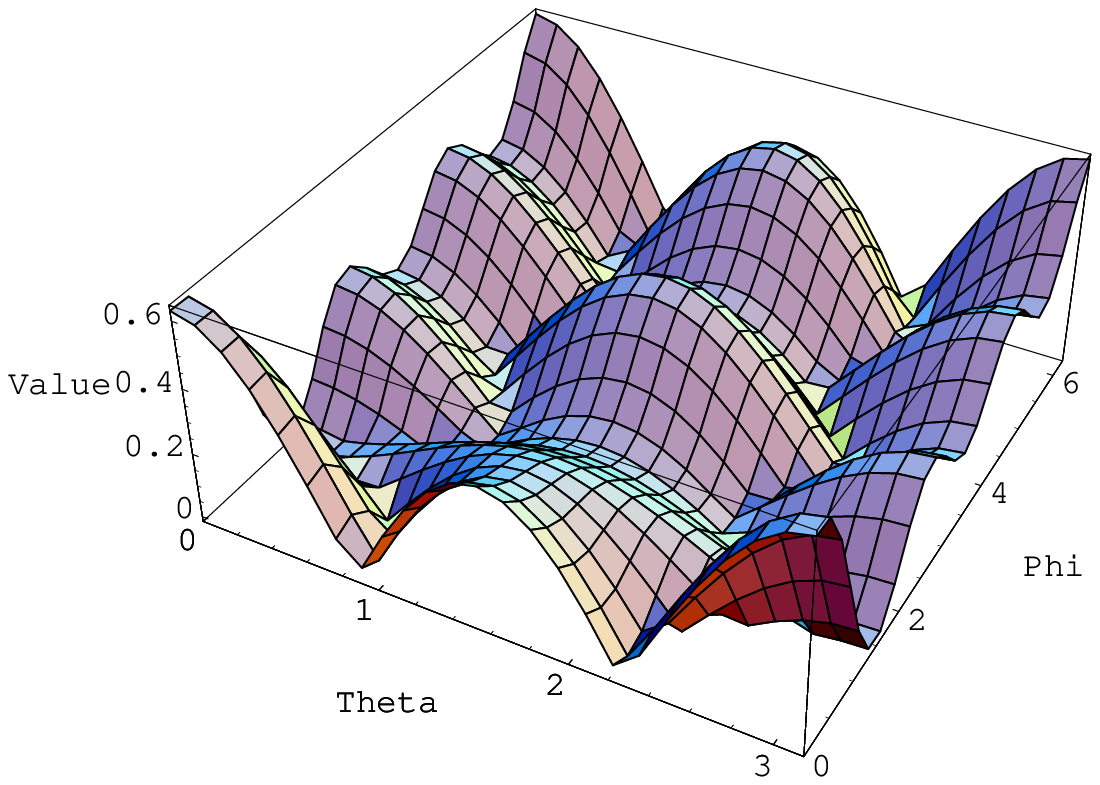}

\caption{The angular dependence to the $+$ polarization for the Virgo interferometer
at 8000 Hz}
\end{figure}

\begin{figure}
\includegraphics{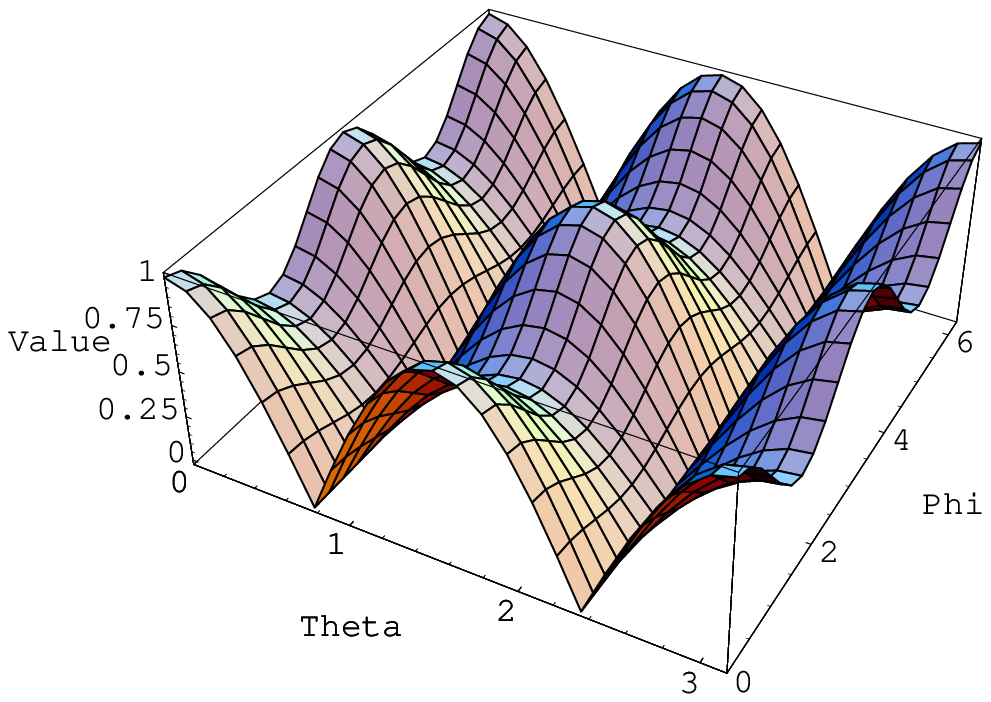}

\caption{The low-frequency angular dependence to the $+$ polarization for
the LIGO interferometer}
\end{figure}

\begin{figure}
\includegraphics{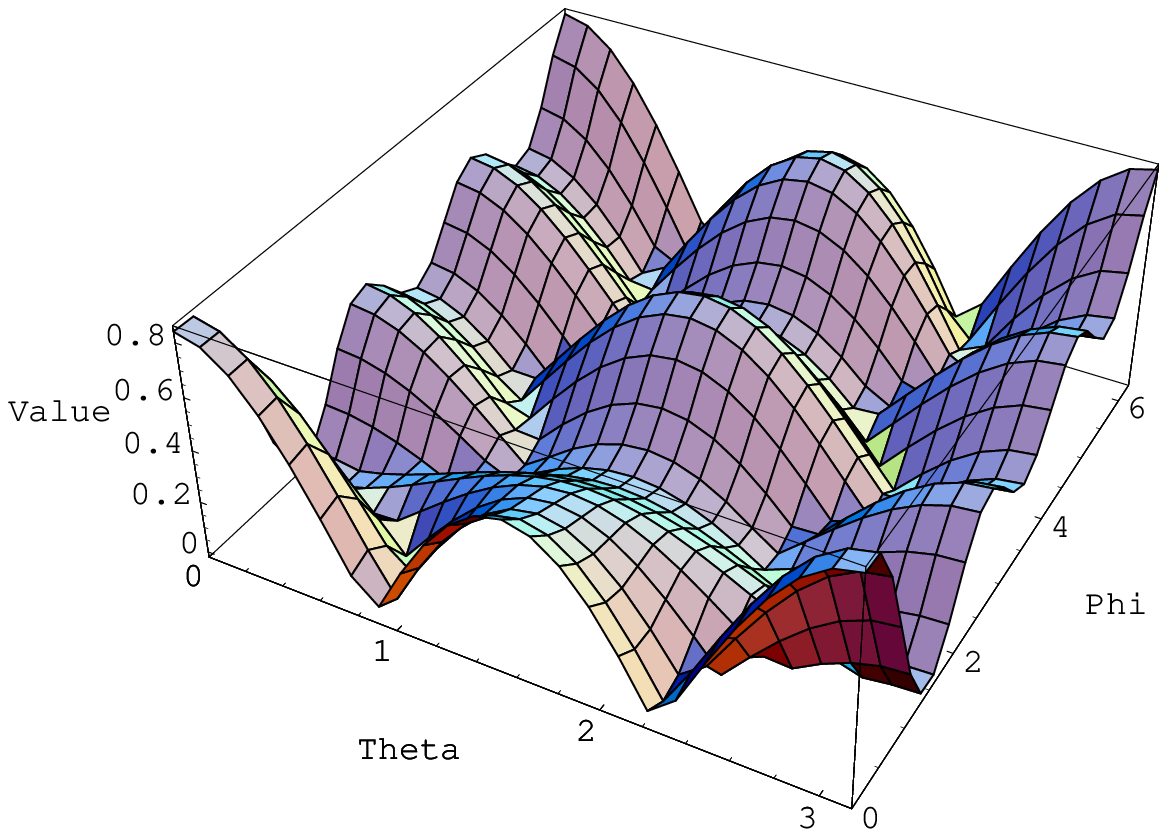}

\caption{The angular dependence to the $+$ polarization for the LIGO interferometer
at 8000 Hz}
\end{figure}

\begin{figure}
\includegraphics{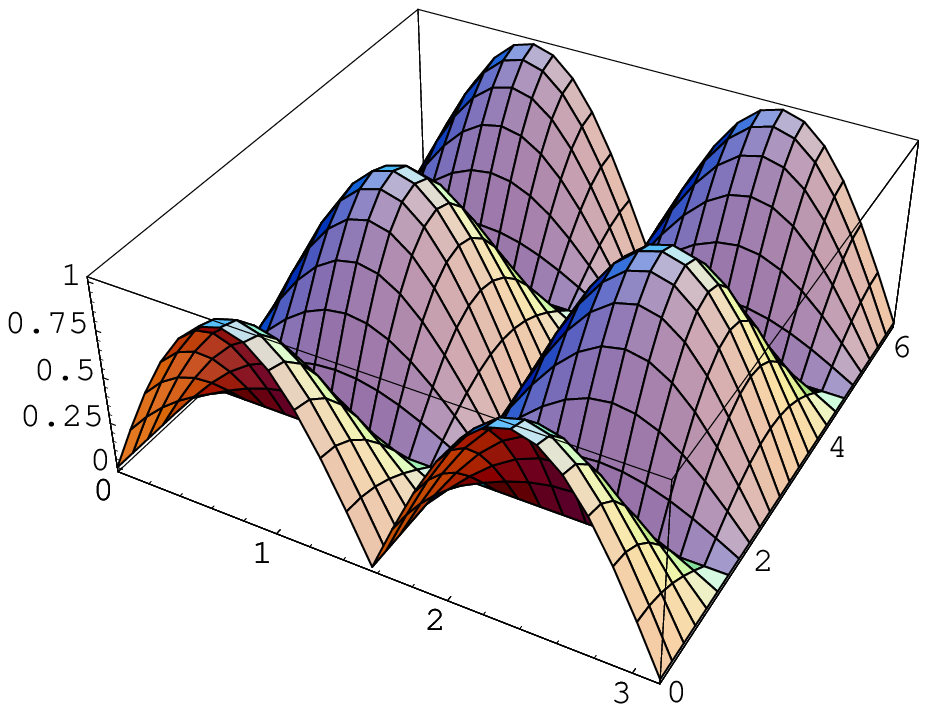}

\caption{The low-frequency angular dependence to the $\times$ polarization
for the Virgo interferometer }
\end{figure}

\begin{figure}
\includegraphics{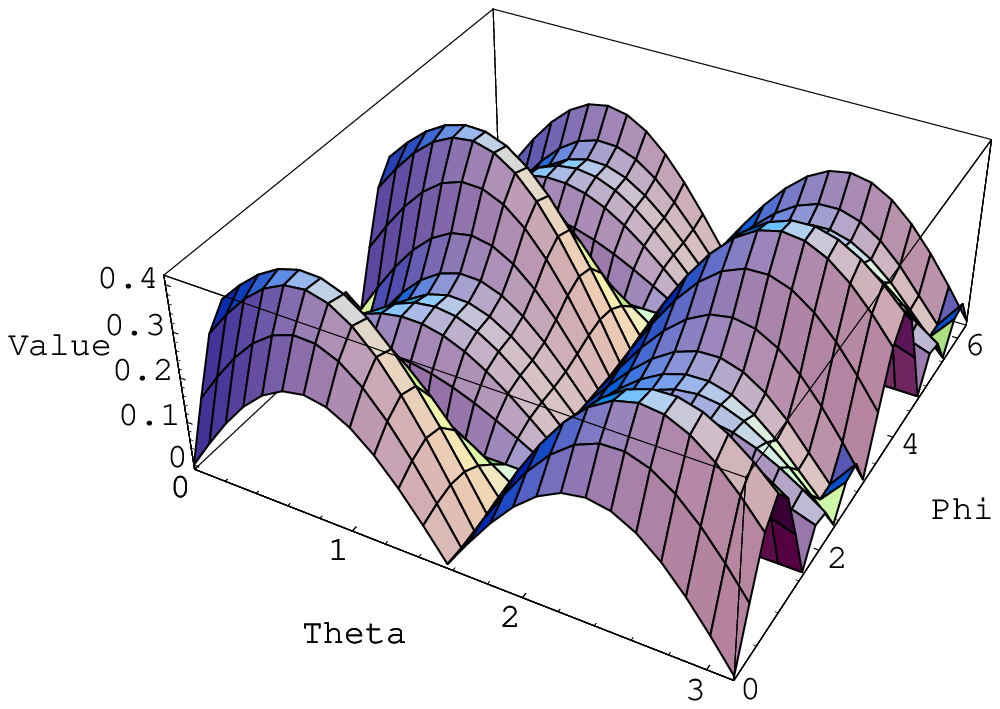}

\caption{The angular dependence to the $\times$ polarization for the Virgo
interferometer at 8000 Hz}
\end{figure}

\begin{figure}
\includegraphics{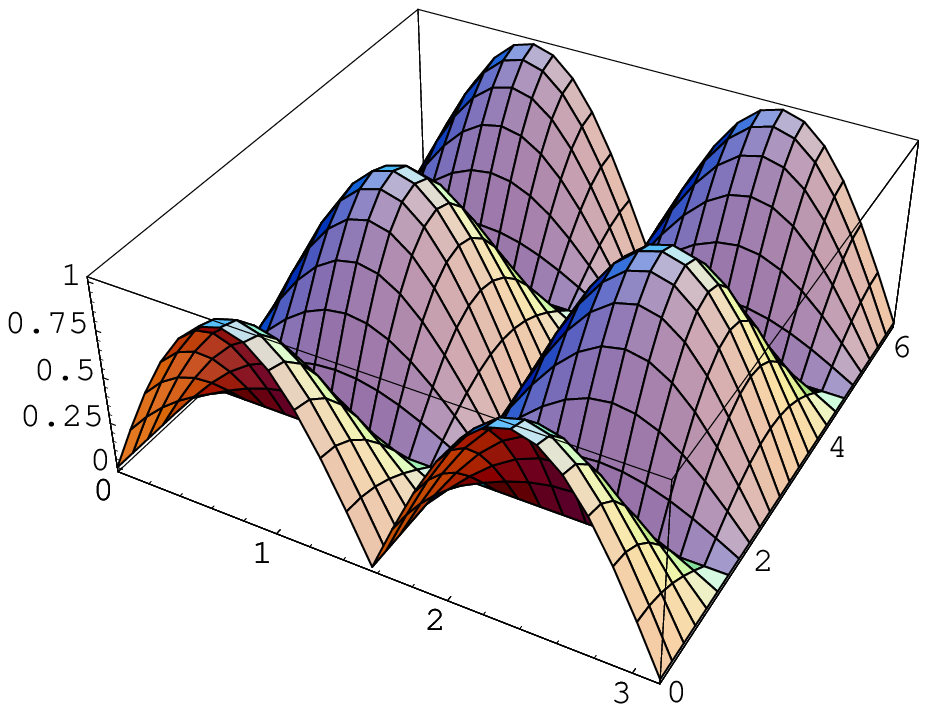}

\caption{The low-frequency angular dependence to the $\times$ polarization
for the LIGO interferometer }
\end{figure}

\begin{figure}
\includegraphics{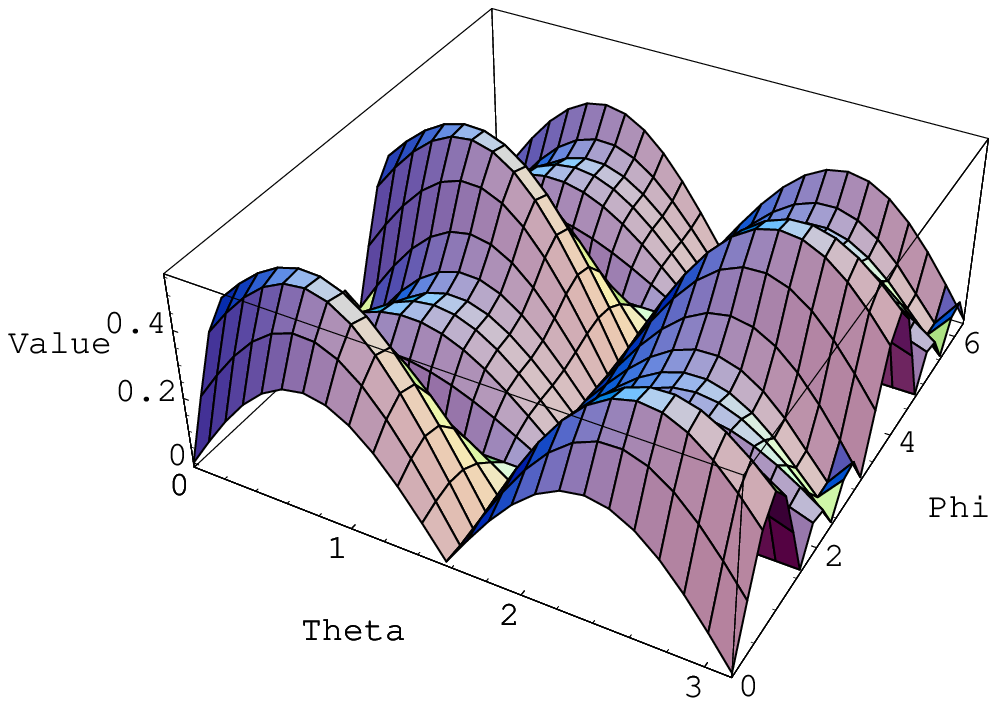}

\caption{The angular dependence to the $\times$ polarization for the LIGO
interferometer at 8000 Hz}
\end{figure}

Seeing the Figures 16 and 18 of eq. (\ref{eq: risposta totale Virgo + full})
and 20 and 22 of (\ref{eq: risposta totale Virgo per full}) at 8000
Hz, one sees that the magnetic component of GWs \textbf{cannot} extend
the frequency range of interferometers. This is because, even if magnetic
contributions grow with frequency, as it is shown from eq. (\ref{eq: risposta totale 2 acc}),
the division between {}``electric'' and {}``magnetic'' contributions
breaks down at high frequencies, thus one has to perform computations
using the full theory of gravitational waves. The correspondent response
functions which are obtained do not grow which frequency.

\section{Conclusions}

Recently, arising from an enlighting analysis of Baskaran and Grishchuk
in \cite{key-24}, some papers in the literature have shown the presence
and importance of the so-called {}``magnetic'' components of GWs,
which have to be taken into account in the context of the total response
functions of interferometers for GWs propagating from arbitrary directions.
In \cite{key-25} and \cite{key-26} accurate response functions for
the Virgo and LIGO interferometers have been analysed. 

However, some results which have been shown in \cite{key-25} look
in contrast with the results which have been shown in\cite{key-26}.
In fact, in \cite{key-25} it was claimed that the {}``magnetic''
component of GWs could, in principle, extend the frequency range of
Earth based interferometers, while in \cite{key-26} such a possibility
has been banned.

This contrast has been partially solved in \cite{key-27}\textit{.}

The aim of this review paper has been to re-analyse all the framework
of the {}``magnetic'' components of GWs with the goal of solving
the mentioned contrast in definitive way.

Accurate response funtions for the Virgo and LIGO interferometers
have been re-discussed in detail too.

\section*{Acknowledgements}

I would like to thank Professor Frank Columbus and Nova Science Publishers
Inc. for commissioning this review.


\begin{thebibliography}{10}
\bibitem{key-1}F. Acernese et al. (the Virgo Collaboration) - Class. Quant. Grav.
\textbf{24,} 19, S381-S388 (2007)
\bibitem{key-2}C. Corda - Astropart. Phys. \textbf{27,} No 6, 539-549 (2007); 
\bibitem{key-3}C. Corda - Int. J. Mod. Phys. D \textbf{16,} 9, 1497-1517  (2007) 
\bibitem{key-4}B. Willke et al. - Class. Quant. Grav. \textbf{23} 8S207-S214 (2006) 
\bibitem{key-5}D. Sigg (for the LIGO Scientific Collaboration) - www.ligo.org/pdf\_public/P050036.pdf
\bibitem{key-6}B. Abbott et al. (the LIGO Scientific Collaboration) - Phys. Rev.
D 72, 042002 (2005) 
\bibitem{key-7}M. Ando and the TAMA Collaboration - Class. Quant. Grav. \textbf{19}
7 1615-1621 (2002)
\bibitem{key-8}D. Tatsumi, Tsunesada Y and the TAMA Collaboration - Class. Quant.
Grav. \textbf{21} 5 S451-S456 (2004) 
\bibitem{key-9}C. Corda - J. Cosmol. Astropart. Phys. JCAP04009 (2007)
\selectlanguage{italian}
\bibitem{key-10}C. Corda - Int. Journ. \foreignlanguage{english}{Mod. Phys. A 23,
10, 1521-1535 (2008)}
\selectlanguage{english}
\bibitem{key-11}G. Allemandi, M. Francaviglia, M. L. Ruggiero and A. Tartaglia - Gen.
Rel. Grav. 37 11 (2005)
\bibitem{key-12}S. Capozziello and C. Corda - Int. J. Mod. Phys. D \textbf{15,} 1119
-1150 (2006) 
\bibitem{key-13}S. Capozziello, M. F. De Laurentis and M. Francaviglia - Astropart.
Phys. \textbf{2,} No 2, 125-129 (2008) 
\bibitem{key-14}C. Corda- Astropart. Phys. 28, 247-250 (2007)
\bibitem{key-15}Allemandi G, Capone M, Capozziello S and Francaviglia M - Gen. Rev.
Grav. 38 1 (2006)
\bibitem{key-16}Capozziello S and Francaviglia M - arXiv:0706.1146, to appear in Gen.
Rel. Grav. (2007)
\bibitem{key-17}M. E. Tobar , T. Suzuki and K. Kuroda \foreignlanguage{italian}{Phys.}
Rev. \foreignlanguage{italian}{D 59 \textbf{}102002 (1999)}
\bibitem{key-18}K. Nakao, T. Harada , M. Shibata, S. Kawamura and T. Nakamura - Phys.
Rev. D 63, 082001 (2001)
\bibitem{key-19}C. Corda and M. F. De Laurentis - Proceedings of the 10th ICATPP Conference
on Astroparticle, Particle, Space Physics, Detectors and Medical Physics
- Applications, Villa Olmo, Como, Italy (October 8-12 2007)
\bibitem{key-20}C. Corda - Mod. Phys. Lett. A No. 22, 16, 1167-1173 (2007)
\bibitem{key-21}S. Capozziello, C. Corda and M. F. De Laurentis - Mod. Phys. Lett.
A 22, 15, 1097-1104 (2007)
\bibitem{key-22}C. Corda - Mod. Phys. Lett. A No. 22, 23, 1727-1735 (2007)
\bibitem{key-23}C. Brans and R. H. Dicke - Phys. Rev. 124, 925 (1961)
\bibitem{key-24}D. Baskaran and L. P. Grishchuk - Class. Quant. Grav. \textbf{21}
4041-4061 (2004) 
\bibitem{key-25}C. Corda - \foreignlanguage{italian}{Int. Journ.} Mod. Phys. A 22,
13, 2361-2381 (2007)
\bibitem{key-26}C. Corda - Int. J. Mod. Phys. D \textbf{16,} 9, 1497-1517  (2007)
\bibitem{key-27}C. Corda - Proceedings of the XLIInd Rencontres de Moriond, Gravitational
Waves and Experimental Gravity, p. 95, Ed. J. Dumarchez and J. T.
Tran, Than Van, THE GIOI Publishers (2007) 
\selectlanguage{italian}
\bibitem{key-28}Misner CW, Thorne KS and Wheeler JA - {}``Gravitation'' - W.H.Feeman
and Company - 1973 \foreignlanguage{english}{}
\selectlanguage{english}
\bibitem{key-29}Tinto M, Estabrook FB and Armstrong JW - Phys. Rev. D \textbf{65}
084003 (2002) 
\bibitem{key-30}Thorne KS - \textit{300 Years of Gravitation} - Ed. Hawking SW and
Israel W Cambridge University Press p. 330 (1987)
\bibitem{key-31}Saulson P - \textit{Fundamental of Interferometric Gravitational Waves
Detectors} - World Scientific, Singapore (1994) 
\end{thebibliography}
\end{document}